\documentclass[onecolumn]{aastex631}
\usepackage{mathtools}
\usepackage{amsmath}
\usepackage{soul}
\usepackage{lipsum}
\usepackage{gensymb}
\usepackage{rotating}

\newcommand\kms{$\mathrm{km\ s^{-1}}$}
\newcommand\Rsun{$\mathrm{R_{\odot}}$}
\newcommand\Rss{$\mathrm{R_{ss}}$}

\newcommand\Ahe{$\mathrm{A_{He}}$}
\newcommand\vap{$\mathrm{\Delta v_{\alpha p}}$}
\newcommand\vapn{$\mathrm{v_{\alpha, p} / v_A}$}
\newcommand\MA{$\mathrm{M_A}$}
\newcommand\MS{$\mathrm{M_S}$}
\newcommand\MMS{$\mathrm{M_{MS}}$}
\newcommand\sigmac{$\mathrm{\sigma_C}$}
\newcommand\sigmar{$\mathrm{\sigma_R}$}
\newcommand\fss{$\mathrm{f_{ss}}$}
\newcommand\zp{$\mathrm{\mathbf{z}^{+}}$}
\newcommand\zm{$\mathrm{\mathbf{z}^{-}}$}
\newcommand\dv{$\mathrm{\delta \mathbf{v}}$}
\newcommand\db{$\mathrm{\delta \mathbf{b}}$}
\newcommand\angvb{$\mathrm{\theta_{VB}^{sc}}$}
\newcommand\alf{Alfv\'en}
\newcommand\alfic{Alfv\'enic}
\newcommand\alfty{Alfv\'enicity}
\newcommand\SO{Solar Orbiter}
\newcommand\elas{Els\"{a}sser}

\newcommand\heliocast{\protect\footnote{\url{https://helioforecast.space/icmecat/ICME_Wind_MOESTL_20230323_01}}}

\newcommand\mascode{\protect\footnote[8]{\url{https://www.predsci.com/mas/}}}

\shorttitle{Near subsonic solar wind at PSP}
\shortauthors{Ervin et al.}

\graphicspath{{./}{figures}}

\begin{document}

\title{Near subsonic solar wind outflow from an active region}

\author[0000-0002-8475-8606]{Tamar Ervin}
\affiliation{Department of Physics, University of California, Berkeley, Berkeley, CA 94720-7300, USA}
\affiliation{Space Sciences Laboratory, University of California, Berkeley, CA 94720-7450, USA}

\author[0000-0002-1989-3596]{Stuart D. Bale}
\affiliation{Department of Physics, University of California, Berkeley, Berkeley, CA 94720-7300, USA}
\affiliation{Space Sciences Laboratory, University of California, Berkeley, CA 94720-7450, USA}

\author[0000-0002-6145-436X]{Samuel T. Badman}
\affiliation{Center for Astrophysics $\vert$ Harvard \& Smithsonian, 60 Garden Street, Cambridge, MA 02138, USA}

\author[0000-0002-4625-3332]{Trevor A. Bowen}
\affiliation{Space Sciences Laboratory, University of California, Berkeley, CA 94720-7450, USA}

\author[0000-0002-1859-456X]{Pete Riley}
\affiliation{Predictive Science Inc., San Diego, CA 92121, USA}

\author[0000-0002-4559-2199]{Kristoff Paulson}
\affiliation{Center for Astrophysics $\vert$ Harvard \& Smithsonian, 60 Garden Street, Cambridge, MA 02138, USA}

\author[0000-0002-8748-2123]{Yeimy J. Rivera} 
\affiliation{Center for Astrophysics $\vert$ Harvard \& Smithsonian, 60 Garden Street, Cambridge, MA 02138, USA}

\author[0000-0002-4559-2199]{Orlando Romeo}
\affiliation{Department of Earth and Planetary Science, University of California, Berkeley, CA 94720, USA}
\affiliation{Space Sciences Laboratory, University of California, Berkeley, CA 94720-7450, USA}

\author[0000-0002-1128-9685]{Nikos Sioulas}
\affiliation{Department of Earth, Planetary, and Space Sciences, University of California, Los Angeles, CA, USA}

\author[0000-0001-5030-6030]{Davin E. Larson}
\affiliation{Space Sciences Laboratory, University of California, Berkeley, CA 94720-7450, USA}

\author[0000-0003-1138-652X]{Jaye L. Verniero}
\affiliation{Goddard Space Flight Center, Greenbelt, MD 20771, USA}

\author[0000-0003-4437-0698]{Ryan M. Dewey} 
\affiliation{Department of Climate and Space Sciences and Engineering, University of Michigan, Ann Arbor, MI 48109, USA}

\author[0000-0002-9954-4707]{Jia Huang}
\affiliation{Space Sciences Laboratory, University of California, Berkeley, CA 94720-7450, USA}
 
\begin{abstract} 
During Parker Solar Probe (Parker) Encounter 15 (E15), we observe an 18-hour period of near subsonic ({\MS}$\sim 1$) and sub-{\alfic} (SA), {\MA}$<<< 1$, slow speed solar wind from 22 to 15.6 {\Rsun}. As the most extreme SA interval measured to date and skirting the solar wind sonic point, it is the deepest Parker has probed into the formation and acceleration region of the solar wind in the corona. The stream is also measured by Wind and MMS near 1AU at times consistent with ballistic propagation of this slow stream. We investigate the stream source, properties and potential coronal heating consequences via combining these observations with coronal modeling and turbulence analysis. Through source mapping, in situ evidence and multi-point arrival time considerations of a candidate CME, we determine the stream is a steady (non-transient), long-lived and approximately Parker spiral aligned and arises from overexpanded field lines mapping back to an active region. Turbulence analysis of the {\elas} variables shows the inertial range scaling of the {\zp} mode ($f \sim ^{-3/2}$) to be dominated by the slab component. We discuss the spectral flattening and difficulties associated with measuring the {\zm} spectra, cautioning against making definitive conclusions from the {\zm} mode. Despite being more extreme than prior sub-{\alfic} intervals, its turbulent nature does not appear to be qualitatively different from previously observed streams. We conclude that this extreme low dynamic pressure solar wind interval (which has the potential for extreme space weather conditions) is a large, steady structure spanning at least to 1AU.


\end{abstract}

\section{Introduction} \label{sec: intro}

Parker Solar Probe (Parker; \citet{Fox-2016}) and Solar Orbiter \citep{Muller-2020} provide revolutionary measurements of the fields and particle properties of the solar wind. These spacecraft, in combination with other instruments at 1AU, allow for an unparalleled look at the characteristics of the solar wind, and when coupled with modeling methods, allow us to study and parameterize the source regions of the wind these spacecraft measure. We are interested in a period of near sub-sonic and sub-{\alfic} (SA) solar wind observed by Parker and throughout the heliosphere. These extreme solar wind conditions are reminiscent of the 1999 \lq{}day the solar wind died\rq{} event \citep{Smith-2001} and have space weather implications for lower solar wind dynamical pressure leading to the expansion of planetary magnetospheres \citep{Halekas-2023}.

From 2023-03-16/11:54:58 to 2023-03-17/05:57:20, Parker observed plasma that was both sub-sonic and SA -- the fast magnetosonic ({\MMS} = $v_R / \sqrt{v_A^2 + c_s^2}$) and {\alf} ({\MA} = $v_R / v_A$) Mach numbers were $\sim$ 0.1, and the sonic ({\MS} = $v_R / c_s$) Mach $\sim$ 1. {\MA}, {\MMS}, and {\MS} are determined such that $v_R$ is the proton radial velocity, $\mathrm{v_A = |\mathbf{B}|/\sqrt{\mu_0 \rho}}$ is the {\alf} speed and $c_s = \sqrt{\gamma_e k_B (T_e + T_p) / m_i}$ is the speed of sound in the corona, with an adiabatic index ($\gamma_e$) of 1.29 determined by \citep{Dakeyo-2022}. While regions of SA plasma have previously been identified in every Parker Encounter since Encounter 8 in April 2021 \citep{Kasper-2021, Zhao-2022, Bandyopadhyay-2022}, this stream shows a more extreme drop in {\MA} than ever before, and for the first time appears to approach the sonic point. There have also been a few SA intervals detected at 1AU \citep{Gosling-1982, Smith-2001, Usmanov-2005, Stansby-2020c}, characterized by a dramatic depletion in particle density and flow speed. There is a potential connection of these low-density, sub-{\alfic} intervals observed at Parker with the rarefaction region of transient events leading to a low-density cavity (private communication with with Jagarlamudi). In December 2022, the Mars Atmospheric and Volatile EvolutioN (MAVEN; \citet{Jakosky-2015}) mission and spacecraft at 1AU observed a \lq{}disappearing solar wind event\rq{}, where a density depleted stream encountered Earth then Mars \citep{Halekas-2023}. The stream was seen at spacecraft with large longitudinal, latitudinal, and radial separation, showing decreases to the {\alf} and magnetosonic Mach number \citep{Halekas-2023} and the potential relation to a single coronal source region \citep{Dewey-inprep}.



Crossing into an interval with {\MA} $\leq 1$ is known as crossing the \lq{}{\alf} surface\rq{} and is likely the region where the active heating and acceleration mechanisms such as resonant wave damping and turbulent dissipation take place \citep{Heinemann-1980, Velli-1993, Kohl-1998, Matthaeus-1999, Cranmer-2000, Cranmer-2007, Hollweg-2002, Kasper-2019, Kasper-2021, Chandran-2021, Zank-2021, Zhao-2022, Badman-2023, Bowen-2023}, however it is important to keep in mind that these can also be due to transient low density coincidences. Within the {\alf} surface, inward propagating {\alf} waves remain tethered to the corona, and thus the turbulence properties of the plasma in SA regions likely exhibit differing properties than the rest of the wind. Turbulence in SA regions has been found to show weaker magnetic compressibility \citep{Bandyopadhyay-2022} and a flattening of the inward propagating mode of the {\elas} variables at high frequencies \citep{Zhao-2022, Zank-2022}. The SA regions observed previously were determined to be dominated by outward-propagating {\alf} waves \citep{Kasper-2021, Zhao-2022, Bandyopadhyay-2022, Zank-2022} with some contribution from outward-propagating fast magnetosonic waves \citep{Zhao-2022}. These studies have shown that SA regions are less turbulent and systematically more anistropic than neighboring super-{\alfic} regions \citep{Bandyopadhyay-2022} The extreme drop in {\MA} during this stream will allow us to explore the role of a large {\alf} speed gradient on wave mode generation and conversion and turbulence properties.


Looking at the turbulence properties of the plasma allows us to study potential heating mechanisms of the solar corona and solar wind \citep{Matthaeus-1999, Zank-2018, Zank-2021}. Connecting the observed plasma to its coronal source region will provide additional insight to the properties and characteristics of the solar wind and allows us to better understand the source properties of streams with extremely low Mach numbers. Through the use of potential field source surface (PFSS) and magnetohydrodynamic (MHD) models, we can connect the plasma observed by the spacecraft to its estimated source surface footpoint, and determine where the wind originated. The fast solar wind (FSW) has been shown to originate from open field lines in coronal hole (CH) regions \citep[and others]{McComas-1998, McComas-2008}, however the source region of the slow solar wind is more elusive. There are various theories for the source region of the slow wind such as: the S-web \citep{Chitta-2023, Lynch-2023}, pseudostreamers \citep[and others]{Eselevich-1999, Riley-2012, Wang-2012, Wang-2019, PV-2013, Owens-2014, Abbo-2015, DAmicis-2021aa, Telloni-2023, Velli-2020, Panasenco-2020}, small CHs and their boundaries \citep{DAmicis-2021}, and active region (AR) contributions, especially during solar maximum \citep{Liewer-2004aa, Harra-2008, Kasper-2016, Alterman-2018, Alterman-2019, Brooks-2020, Stansby-2021aa}. Outflows from the edges of ARs are likely caused by reconnection of over-expanded coronal loops above the AR or the opening of these loops to interplanetary space \citep{Harra-2008}.  Typically, plasma originating in CH regions shows high alpha-to-proton abundance ratios \citep{Ohmi-2004}, large differential velocities \citep{Stansby-2020comp}, and is highly {\alfic} \citep{DAmicis-2015, Perrone-2020, DAmicis-2021}. In contrast to this, plasma from ARs have low alpha-to-proton abundance ratios, low differential velocities, high charge state ratios \citep{Liewer-2004aa}, and varying levels of {\alfty}. As shown later, we find this stream to be continuous outflow from an active region. While the contributions of coronal holes to solar wind outflow are relatively well understood, the picture around solar wind from ARs is much more complex. The relationship to SA flows at 1AU has been hinted at but have never been probed this close to the Sun.


In this paper, we study the 18-hour near sub-sonic and SA interval observed by Parker during Encounter 15 (E15 hereafter) to understand the stream's origin, evolution, and turbulence properties. In Section~\ref{sec: data}, we give an overview of the plasma parameters of this stream at Parker. We discuss multi-spacecraft observations of the stream in Section~\ref{sec: multisc} to look at the effect of radial propagation on its properties. In Section~\ref{sec: turbulence} we discuss the analysis of turbulence properties for the stream and compare our results with typical results in the \lq{}ambient\rq{} solar wind. Section~\ref{sec: source-region} discusses the relation to transients and the modeling methods used to connect this stream to an active region source. We outline our results and conclusions and implications in Sections~\ref{sec: results} and \ref{sec: conclusion}. In Appendix~\ref{sec: appendix-model-val} we give an overview of model validation methods and Appendix~\ref{sec: appendix-turb} discusses the calculation of turbulence spectra and the quantification of uncertainty related to the finite velocity grid of the SPAN-I instrument.

\section{Observations} \label{sec: data}

Since launching in 2018, Parker has completed 17 of its 24 planned orbits, reaching its closest heliocentric distance to date (11.4{\Rsun}) on September 27, 2023 during Encounter 17. The mission consists of four instruments taking in situ and remote measurements to study energy transport in the corona and the properties of the solar wind. We use DC magnetic field measurements from the Electromagnetic Fields Investigation (FIELDS; \citet{Bale-2016}) fluxgate magnetometer to study the structure and compressibility of the magnetic field. Proton and alpha particle density, velocity, and temperature measurements measured by the ion Solar Probe ANalyzer (SPAN-I; \citet{Livi-2022}) are used alongside FIELDS measurements to determine pressure, Mach numbers, and turbulence parameters. Electron density and temperature is calculated from a quasi-thermal noise (QTN) fit following the methods of \citet{Romeo-2023} from the SPAN-E 3D electron velocity distribution functions \citep{Whittlesey-2020}.


Since launch, Parker Solar Probe has seen multiple Parker spiral alignments, periods where both spacecraft are measuring the same footpoint longitude in Carrington coordinates, with inner heliosphere and 1AU instruments. These periods allows us to use multi-spacecraft observations to determine the effects of evolution on the bulk solar wind properties of a stream \citep{Rivera-2023, Ervin-2024SSW}. During the SA stream, Parker is in alignment with {\SO} \citep{Muller-2020}, Wind, and the Magnetosonic Multiscale (MMS) mission \citep{Burch-2016}. From {\SO}, we use magnetic field measurements from the Magnetometer (MAG; \citet{Horbury-2020}) and proton density, temperature, and velocity from the Solar Wind Analyzer (SWA; \citet{Owen-2020}) to look at evolutionary effect in the inner heliosphere. The complimentary observations on Wind are: DC magnetic field measurements from the Wind Magnetic Field Investigation (MFI; \citet{Lepping-1995}) and proton and electron bulk properties from the Wind Three-Dimensional Plasma and Energetic Particle Investigation (3DP; \citet{Lin-1995aa}) to study the solar wind out to 1AU. The MMS spacecraft each include identical scientific instruments, one of which is the hot plasma composition analyzer (HPCA; \citet{Young-2016}) which studies the plasma properties of minor ion species through resolving the plasma velocity distribution functions (VDFs). During its orbit, MMS crosses the bow shock into the ambient solar wind, providing an additional viewpoint into the processes governing solar wind composition. Using the trajectory of the spacecraft, we identify the SA stream of interest in the MMS data during a region where it was sampling solar wind plasma.

We use modeling methods to map the observed plasma back to an estimated source surface footpoint. These methods rely upon an observed photospheric radial magnetic field as an input. We use full disk magnetograms from the Global Oscillation Network Group (GONG; \citet{Harvey-1996}) with the Air Force Data Assimilative Photospheric Flux Transport (ADAPT; \citet{Worden-2000, Arge-2010, Arge-2011, Arge-2013, Hickmann-2015}) model applied. This model creates a more accurate full-Sun Carrington magnetic field map by modeling the effects of far side evolution on the photospheric radial magnetic field. In addition, we use full disk images from the 193{\AA} bandpass filter on the Atmospheric Imaging Assembly (AIA; \citet{Lemen-2012}) aboard the Solar Dynamics Observatory (SDO; \citet{Pesnell-2012}) to create full-Sun EUV maps for comparison with our modeling results. SDO/AIA consists of four telescopes taking images of the solar disk at a 12 second cadence, which allows us to create a high resolution full-Sun map to estimate the footpoint brightness and provide visual comparisons with modeling results. 


\subsection{Plasma Parameters} \label{sec: data-params}

\begin{figure} [htb!]
  \includegraphics[width=\columnwidth]{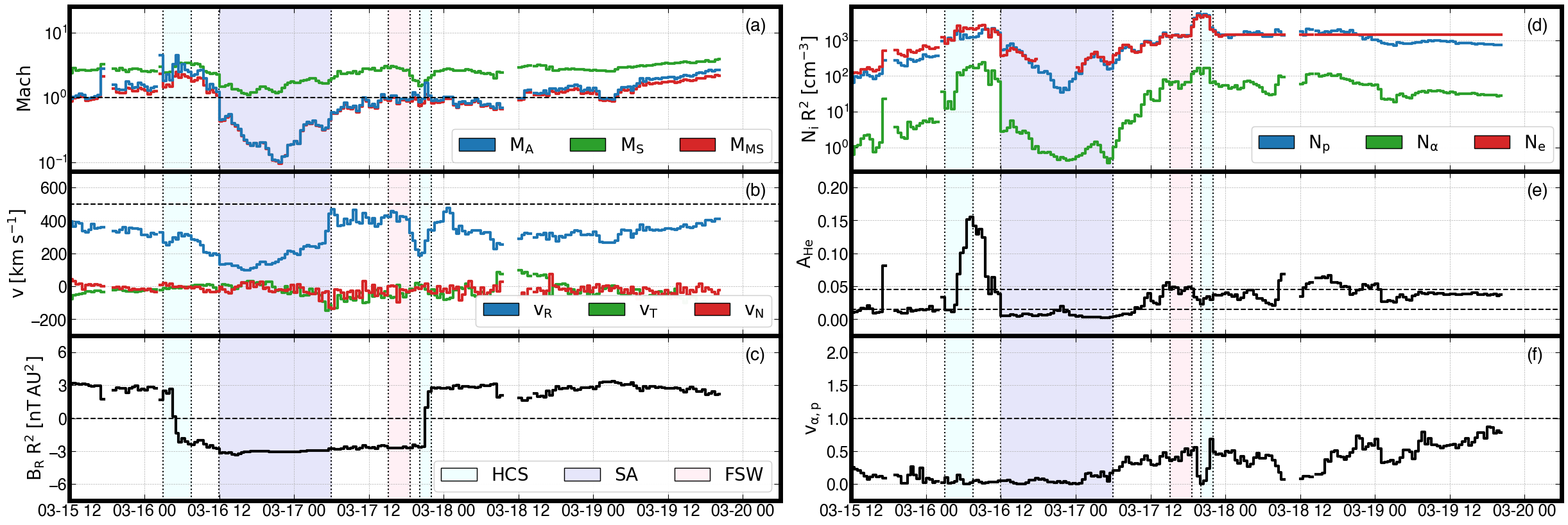}
  \caption{Overview of in situ plasma parameters at Parker from 03/15 to 03/20 showing identification of the streams of interest: the SA region (purple), HCS crossings (blue), and fast wind (pink). We identify the fast wind stream to use as a validation period. All observations are shown at a 30-minute cadence.
  \textit{Panel (a):} {\alf} (blue), sonic (green) and magnetosonic (red) Mach numbers calculated from Parker/FIELDS and SWEAP measurements. The dashed line at 1 shows the cutoff where the wind speed is equal to the Mach number. The y-axis is in a log scale.
  \textit{Panel (b):} Proton radial (blue), tangential (green), and normal (red) velocity as measured by Parker/SWEAP. The dashed line shows the canonical 500 {\kms} cutoff between fast and slow wind at 1AU. 
  \textit{Panel (c):} The scaled radial magnetic field as measured by Parker/FIELDS. The dashed line at 0 shows the crossing of the neutral line by the spacecraft. 
  \textit{Panel (d):} The scaled proton, alpha particle, electron densities. Proton and alpha densities come from Parker/SWEAP, and the electron density is calculated from QTN following the methods of \citet{Romeo-2023}. The gap in electron density during the SA region is due to issues with the QTN analysis during this period.
  \textit{Panel (e):} Alpha-to-proton abundance ratio from Parker/SWEAP. The dashed lines at 0.015 and 0.045 show the cutoffs for low and high alpha abundance ratios \citep{Kasper-2007, Kasper-2012}.
  \textit{Panel (f):} Alpha-to-proton differential velocity ({\vap} $\mathrm{= v_{\alpha} - v_p}$) normalized by the {\alf} speed ($\mathrm{v_A}$). The dashed line at 1 shows where the differential velocity is equal to the {\alf} speed.
  }
  \label{fig: identification}
\end{figure}

In Figure~\ref{fig: identification}, we show an overview of the plasma properties as seen by Parker from March 15 to March 20. We identify the SA stream, spanning from 11:54:58 on March 16 (22 {\Rsun}) to 05:57:20 on March 17 (15.6 {\Rsun}), in purple and identify other features of interest for added context. The two heliospheric current sheet (HCS) crossings, early March 16 and late March 17, are highlighted in blue, and provide a method to validate our magnetic field models and cross reference observations at different spacecraft. The FSW stream (highlighted in pink) is used as another validation of our methods and spans from 13:52 to 18:37 on March 17.

We identify an 18 hour extremely SA, near subsonic, period by using Parker observations (proton density and magnetic field) to calculate the {\alf} ({\MA}), sonic ({\MS}), and fast magnetosonic ({\MMS}) Mach numbers (see panel (a) in Figure~\ref{fig: identification}). A period is SA when {\MA} $\mathrm{= v_R/v_A}$$\leq 1$, meaning that the solar wind at this point is slower than the {\alf} speed ($\mathrm{v_A = |\mathbf{B}|/\sqrt{\mu_0 \rho}}$). We calculate {\MS} = $v_R / c_s$ and {\MMS} $\mathrm{= v_R/ \sqrt{v_A^2 + c_s^2}}$, with a sound speed defined as $c_s = \sqrt{\gamma_e k_B (T_e + T_p) / m_i}$, such that $T_e$ is the electron temperature fit \citep{Romeo-2023}, $T_p$ is the proton temperature from SPAN-I, $m_i$ is the ion mass, and $\gamma_e$ is the adiabatic index of 1.29 as defined by \citet{Dakeyo-2022}. During this period, the {\MA} and fast magnetosonic Mach ({\MMS}) drops to extremely low values ($\sim 0.1$) and the sonic Mach ({\MS}) is $\sim 1$. Figure~\ref{fig: identification} includes a data gap during the SA period in the $N_e$ observations due to unphysical density values from the QTN analysis.



The SA region (identified in purple) shows in situ characteristics that are dramatically different than the neighboring FSW stream (pink). The SA solar wind period hits some of the lowest speeds observed to date ($\sim 150$ {\kms}) and shows a sharp transition into the region (drop in {\MA} and wind speed) then a gradual return to normal streaming speeds and super-{\alfic} wind. This is contrasted by the FSW stream with velocities $\sim 500$ {\kms} and {\MA} $\sim 1$. In panel (c) we show the scaled radial magnetic field to identify the HCS crossings (shown in blue). The SA stream is in a negative polarity interval located between, but well away from (in longitude), two HCS crossings (panel (c)). This shows we move rapidly in longitude through the SA interval.

The particle densities and alpha-to-proton abundance ratio ({\Ahe} = $\mathrm{N_{\alpha}/N_p}$), share information about the source region of the stream. {\Ahe} varies from $\sim$0 to $\sim$0.20 during E15, showing times of both high (above 0.145) and low (below 0.045) alpha particle abundance indicating a variety of coronal source origins for the plasma measured during this encounter \citep{Kasper-2007, Kasper-2012}. The electron, proton, and alpha particle densities are all depleted during this period (panel (d)) and {\Ahe} $\sim 0$ (panel (e)), which is characteristic of wind emerging from streamers and active regions \citep{Borrini-1981, Gosling-1981, Suess-2009, Kasper-2007, Kasper-2012, Kasper-2016, Alterman-2018, Alterman-2019}. This is a very different in situ signature than those at the HCS and FSW regions. Similar to previous observations, there are enhancements in the proton density at the HCS crossings \citep{Ervin-2024SSW}, yet huge variance in alpha particle abundance between the two crossings \citep{Zhao-2017, Rivera-2021}. The inconsistencies in the alpha particle abundance ratio and heavy ion composition in general is likely due to \lq{}heavy ion dropouts\rq{} in the \lq{}Outlier\rq{} wind \citep{Zhao-2017, Rivera-2021}. Modeling by \citet{Raymond-2022} has shown these dropouts to correlate with dense gas blobs with reduced heating due to local variations in Alfven speed changing the reflection of Alfven waves and the turbulence cascade. The environment in which this happens could impact the number of escaping helium ions leading to modulation of the observed alpha particle abundance in the heliosphere, especially during HCS crossings \citep{Rivera-2021}. This  huge spike in alpha particle abundance at the first crossing is unique and should be explored further potentially as an indicator of leakage at the HCS. The high {\Ahe} ($\sim 0.05$) in the FSW region indicates different coronal origin than for the SA stream, and is typical for wind coming from CHs \citep{Ohmi-2004, Stansby-2020comp}. 


In panel (f) we look at alpha-proton differential streaming to look at particle instabilities in the plasma and how their boundaries transition between SA and super-{\alfic} regions. We calculate the alpha-to-proton differential velocity following the methods of \citet{Reisenfeld-2001, Fu-2018, Huang-2023}, where {\vap} $\mathrm{= (v_{\alpha} - v_p)/ \cos{\theta}}$ such that $v_{\alpha}$ and $v_p$ are the radial velocity of the alpha particles and protons respectively, and $\cos{\theta} = |B_R/B|$. This calculation method removes the dependence on the polarity of the magnetic field. {\vap} is normalized by the {\alf} speed to become {\vapn}, and remains below 1 for the entirety of this encounter meaning that the differential velocity ({\vap}) is less than the {\alf} speed ($\mathrm{v_A}$). This is an expected instability threshold due to wave particle interactions driven by alpha-proton instabilities \citep{Gary-2000}. The SA period shows very different streaming behavior ({\vapn} $\sim 0$), than the FSW period ({\vapn} $\sim 0.5$) which could indicate preferential generation of certain wave modes. The normalized differential velocity is also expected to show some correlation with solar wind speed \citep{Mostafavi-2022}, and during the SA period, we see the lowest normalized differential velocity of the period, reaching a ratio of $\sim$0 in both the regular cadence and binned data for a variety of bin sizes (e.g. 10, 20, 30, and 60-minute bins).

\subsection{Pressure Variations} \label{sec: data-pressure}
\begin{figure} [htb!]
  \includegraphics[width=\columnwidth]{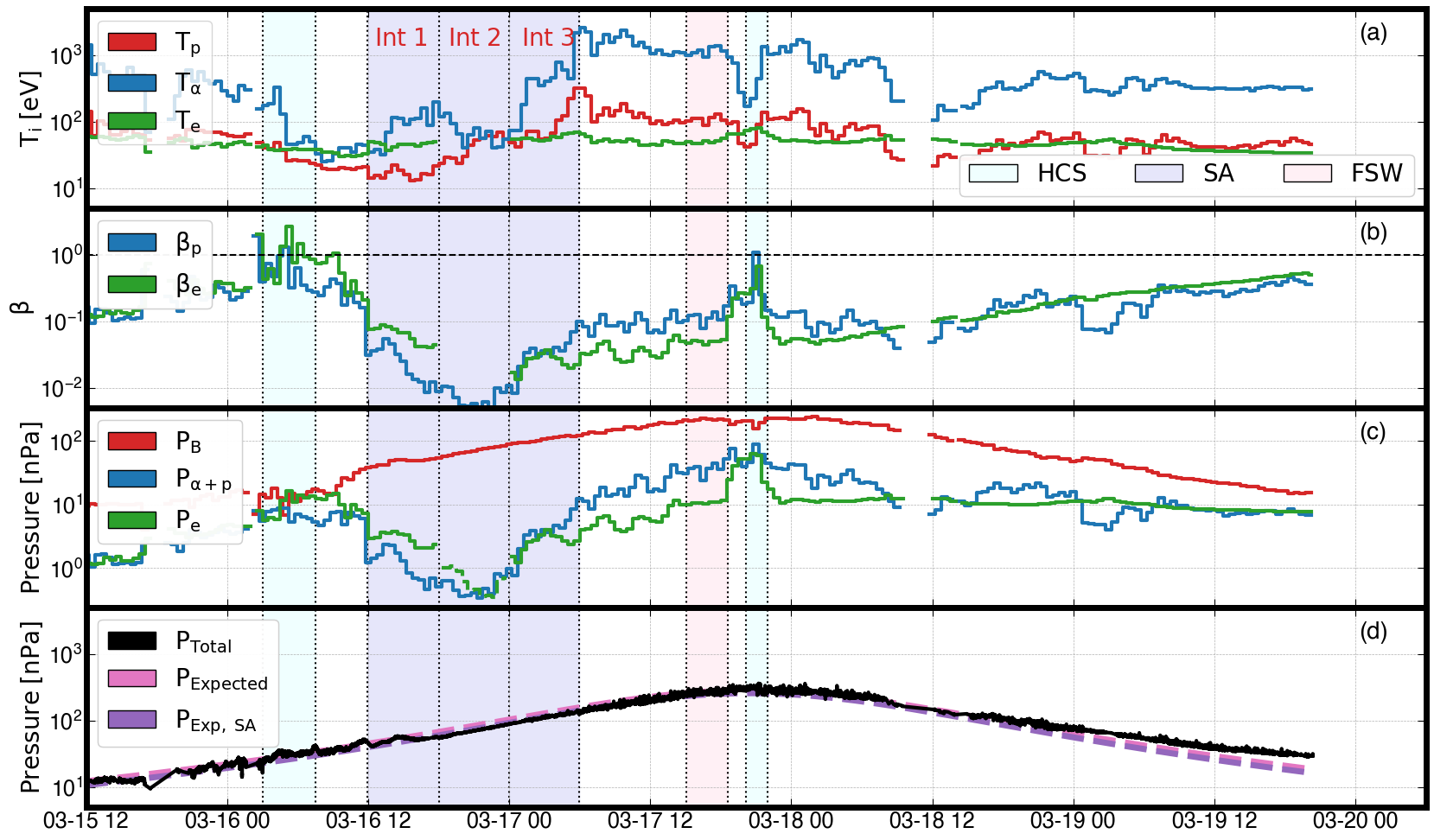}
  \caption{A timeseries of the magnetic, kinetic, and total pressure along with plasma beta for E15. Streams of interest are identified following the convention of Figure~\ref{fig: identification}.
  \textit{Panel (a):} Proton (blue), alpha particle (green), and electron (red) in situ temperatures from SPAN-I and SPAN-E.
  \textit{Panel (b):} Plasma beta for protons (blue) and electrons (red).
  \textit{Panel (c):} Magnetic (red), ion (blue), and electron (green) pressure (nPa). The dashed green line in electron pressure during the SA region is the period where $N_p$ was substituted for $N_e$ in the pressure calculation due to issues with QTN. 
  \textit{Panel (d):} Comparison of the total pressure (black) and expected pressure red. The expected pressure is a function of radial distance and defined by $P_{expt} = 10^{6.61} \times R^{-3.71}$. We fit the total pressure for the SA region (purple dashed line) and find the following:  $P_{expt} = 10^{6.61} \times R^{-3.74}$.
  }
  \label{fig: pressure}
\end{figure}

In Figure~\ref{fig: pressure}, we show the magnetic and kinetic pressure, total pressure, plasma beta ($\beta$) calculated from the Parker measurements shown in Figure~\ref{fig: identification}. Plasma $\beta$ measures the degree of non-uniformity held by the magnetic field for a plasma in equilibrium and probing different $\beta$ regimes allow us to understand various plasma processes that are dependent on the dynamics of the plasma. Low $\beta$ regimes are dominated by the interplay of magnetic forces in balance with each other, $\beta \sim 1$ is pressure balanced, and in $\beta >>> 1$ plasma the magnetic field is not dynamically important. Similarly, studying pressures (panels (c) and (d)), especially at the boundaries of different regions, provide insight to the stability of a stream, degree of adiabatic expansion in our plasma, and highlight the radial dependence of pressure in the plasma.


In panel (b), we show $\beta$ the ratio of the kinetic pressure to magnetic pressure, $\beta = \frac{P_{kinetic}}{P_{magnetic}} = \frac{n k_B T}{|\mathbf{B}|^2 / 2\mu_0}$. $P_{K, species} = n_s k_B T_s$ and $P_{magnetic} = P_B = |\mathbf{B}|^2 / 2\mu_0$. During the SA period, we see extremely low $\beta$ values ($\sim 0.01$) driven by a drop in thermal pressure (panel (c)). This is an order of magnitude deviation from typical values in the solar wind ($\beta \sim 0.1$) such as those seen in the FSW period, and shows very different behavior than the $\beta \sim 1$ HCS wind.

In panel (c) we show the kinetic ($P_e, P_p, P_{\alpha}$ are the pressure of the electrons, protons, and alpha particles) and magnetic ($P_B$) pressures as defined above. As with the parameters in Figure~\ref{fig: identification}, during the SA period, we see a sharp drop entering the stream and the electron pressure is dominant over the ion pressure, then a gradual return to \lq{}normal\rq{} conditions and ion pressure begins to dominate. There is previous evidence that electron pressure is dominant over proton pressure in the acceleration of the slow wind \citep{Dakeyo-2022} and these observations could indicate this as well. Due to issues with the QTN analysis leading to unreliable electron density in interval two of the SA period, we substitute the SPAN-I proton density for electron density to calculate the electron pressure during this period such that we can calculate the total pressure (panel (d)) in the SA period. The affected portion is shown with a dashed line in panel (c) of Figure~\ref{fig: pressure}.

The total pressure (panel (d)) is the sum of the kinetic and magnetic pressures: $\mathrm{P_{total} = P_K + P_B}$ and follows a theoretical relation with radial distance from the solar surface based on the radial scaling of the kinetic and magnetic components. In an idealized adiabatic spherical expansion, the magnetic field and plasma density scales as $1/R^2$. In the regime of the wind that Parker measures, the effective polytropic index ($\gamma$) varies with solar wind type and falls somewhere between isothermal ($\gamma = 1$) and adiabatic ($\gamma = 5/3$) values \citep{Shi-2022, Dakeyo-2022}. For fast wind, adiabatic expansion better fits in situ parameters while slower wind is closer to isothermal. In this case, the temperature scales as $T \propto n^{\gamma - 1} \propto R^{-4/3}$ so the kinetic pressure scales as follows: $P_K \propto n T \propto n n^{\gamma -1} \propto n^{\gamma} \propto R^{-10/3}$. The magnetic pressure scales as $B^2$ so we get $P_B \propto B^2 \propto R^{-4}$. As our total pressure is a sum of the magnetic and kinetic pressures, we can expect the total pressure to have a scaling factor between -4 and -10/3. 

Similar to \citet{Huang-2023}, we fit the total measured pressure to determine an expected total pressure scaling law of the form of $P_{expt} = 10^{a} \times R^{b}$. We find that for the total pressure, $a = 6.61$ and $b = -3.71$, so we have $P_{expt} = 10^{6.61} \times R^{-3.71}$. For the SA region, we have $P_{expt} = 10^{6.61} \times R^{-3.74}$. While the SA period ($\propto R^{-3.74}$) fit does not deviate much from the global fit ($\propto R^{-3.71}$), panels (b) and (c) show that this is entirely driven by magnetic pressure and thus the large variation in thermal pressure is unimportant for the radial scaling of this period. 

\subsection{Multi-point Observations} \label{sec: multisc}

\begin{figure} [htb!]
  \includegraphics[width=\columnwidth]{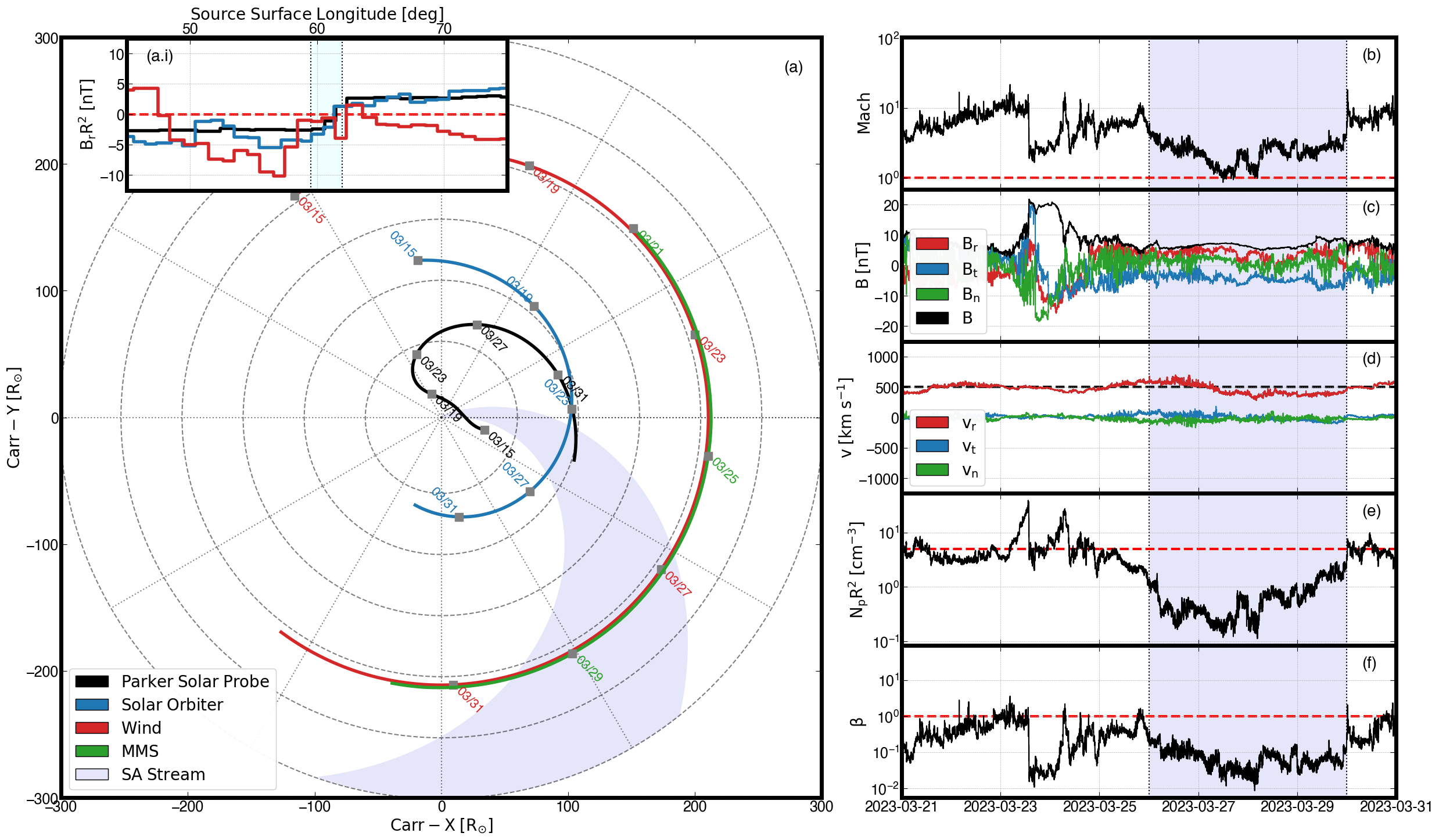}
  \caption{The SA stream region is highlighted in purple in all panels. 
    \textit{Panel (a):} The trajectory of the spacecrafts in Carrington coordinates. The purple shaded region highlights the propagation of the SA stream and its intersection with the spacecraft trajectories.
    \textit{Panel (a.i):} Inset axes shows the scaled radial magnetic field for Parker, {\SO}, and Wind as a function of source surface longitude. The shaded region shows the neutral line crossing alignment.
    \textit{Panel (b):} {\alf} Mach number ($\mathrm{M_A = v_R/v_A}$) with a dashed line at 1 showing where $\mathrm{v_R = v_A}$. 
    \textit{Panel (c):} Magnetic field observations from Wind/MFI. 
    \textit{Panel (d):} The proton radial velocity from Wind/3DP. The dashed line at 500{\kms} shows the canonical 1AU cutoff between fast and slow wind.
    \textit{Panel (e):} The scaled proton number density ($\mathrm{cm^{-3}}$) from Wind/3DP. The dashed line at 5$\mathrm{cm^{-3}}$ shows the ambient 1AU solar wind density.
    \textit{Panel (f):} Plasma $\beta$ calculated from particle density, temperature, and magnetic field measurements. A dashed line at 1 shows where the magnetic pressure balances the plasma pressure. 
  }
  \label{fig: radial}
\end{figure}

\begin{figure} [htb!]
  \includegraphics[width=\columnwidth]{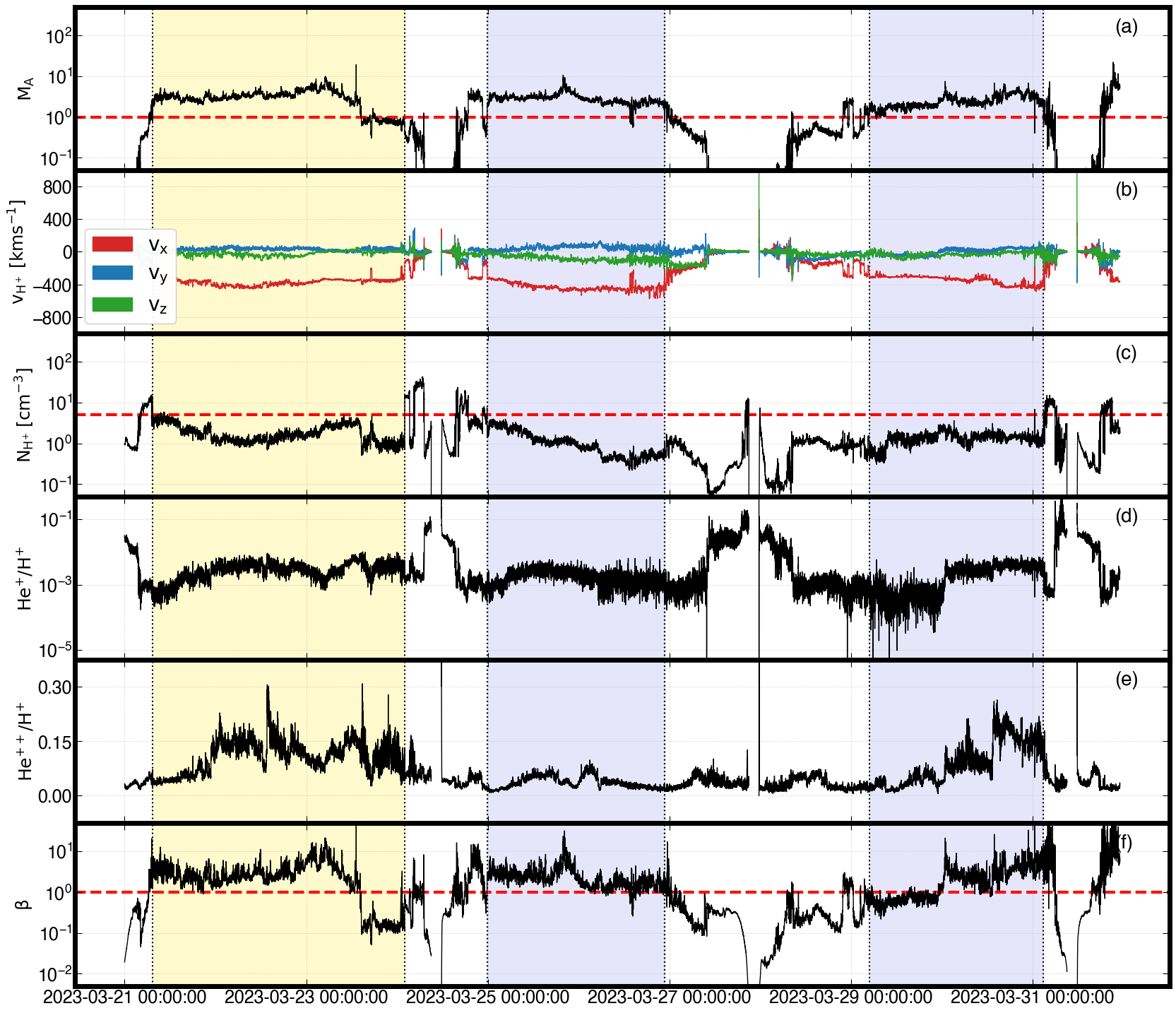}
  \caption{Measurements of particle density and velocity from the HPCA instrument on MMS. The SA region of interest is shown in purple. Other regions where MMS is outside of Earth's magnetosphere and sampling the solar wind are highlighted in yellow.
  \textit{Panel (a):} {\MA} as determined by MMS observations. The dashed line at 1 shows the crossing into sub-{\alfic} plasma.
  \textit{Panel (b):} $\mathrm{H^+}$ velocity measurement.
  \textit{Panel (c):} $\mathrm{H^+}$ density measurement. Typical 1AU $\mathrm{H^+}$ density ($\sim$ 4 $\mathrm{cm}^{-3}$) is shown with a red dashed line.
  \textit{Panel (d):} $\mathrm{He^+/H^+}$ density measurement.
  \textit{Panel (e):} $\mathrm{He^{++}/H^+}$ density measurement.
\textit{Panel (f):} Plasma beta for $\mathrm{H^+}$. The dashed line at 1 shows where the magnetic and kinetic pressure components are balanced.
  }
  \label{fig: mms}
\end{figure}

E15 provides a unique opportunity to study the effects of propagation on the SA stream, probing its stability, radial evolution, and possible relation to transients. Due to the Parker spiral alignment between Parker, {\SO}, and spacecraft at 1AU during portions of the encounter we look at particle velocity, density, and magnetic field measurements from different radial locations in the heliosphere. In Figure~\ref{fig: radial}, we show an overview of the Parker spiral configuration between the spacecraft highlighting the propagation of this stream from a single source region along the basic plasma parameters observed by Wind at 1AU and MMS (Figure~\ref{fig: mms}). We use ballistic propagation with a varying wind speed (the in situ $\mathrm{v_R}$ measurement) to transform the timeseries into measurements as a function of longitude \citep{Snyder-1966, Nolte-1973, Macneil-2022, Koukras-2022, Badman-2020}, and check alignment of the spacecraft neutral line crossings as validation of our propagation methods. {\SO} did not have the SWA or MAG turned on during the SA period we would expect to see the stream, and thus we do not have any {\SO} observations.

Similar trends in the large scale structure of the observables and the strong alignment (within 1{\degree}) between the spacecraft HCS crossings suggests confidence in our back mapping methods and in the Parker spiral alignment of the spacecraft. Additionally, as discussed in Appendix~\ref{sec: appendix-model-val}, the modeled HCS from PFSS and MHD models, aligns well with the observed polarity reversal in both Parker and {\SO} data. Based on an idealized Parker spiral, the expected times for observation of the SA, low density stream at Parker, {\SO}, Wind, and MMS is shown in Figure~\ref{fig: radial}. 

From Wind observations, the {\alf} Mach number almost reaching 1 during this period, and a dramatic density depletion from typical ambient density (Figure~\ref{fig: radial}). Plasma $\beta$ is quite low for this period ($\sim 10^{-1}$), indicating magnetically dominated plasma like at Parker. A few days prior, Wind observed signatures of enhanced magnetic field and a drop in plasma beta associated with a CME on March 23, 2024 at 05:49UT as identified by the Helio4Cast living catalogue{\heliocast} \citep{Mostl-2017, Mostl-2020}.


\begin{table}
  \centering
  \begin{tabular}{|c||c|c|c|c|c|}
    \hline 
    \textbf{Spacecraft} & \textbf{Expected Time} & \textbf{Observed Time} & \textbf{Radial Distance} \\
    \hline
    \hline
    \textbf{Parker} & 03/16 - 03/17 & 03/16 - 03/17  & 15.6{\Rsun} to 22{\Rsun} \\
    \hline
    \textbf{Solar Orbiter} & 03/24 - 03/26 & No Observations & N/A \\
    \hline
    \textbf{Wind} & 03/27 - 03/30 & 03/25 - 03/30 & $\sim$210{\Rsun}  \\
    \hline
    \textbf{MMS} & 03/26 - 03/30 & 03/25 - 03/27 \& 03/29 - 03/31 & $\sim$214{\Rsun} \\
    \hline
  \end{tabular}
  \caption{Breakdown of the expected and observed times and positions of each spacecraft encountering this SA stream. Expected is the time the stream is expected to interact with the spacecraft based on the ballistic propagation shown in Figure~\ref{fig: radial}. Actual is the time when the stream was observed in situ.}
  \label{tab: times}
\end{table}

The composition of solar wind plasma provides additional insight to the properties of the source region and heating mechanisms at work in the plasma. The HPCA instrument aboard MMS allows for the study of the composition of the SA stream of interest, as MMS observed this stream during its time in the ambient solar wind. We identify the regions where MMS is outside the magnetosphere and within the solar wind based on the trajectory of the spacecraft, then use ballistic propagation to determine when MMS is within the source surface longitude corresponding to the stream of interest. In a future study, we will discuss the radial propagation of this stream and the composition metrics as measured by MMS during this stream and the surrounding regions. In Figure~\ref{fig: mms}, the $v_x$ velocity during the SA stream (panel (b)) is much slower in magnitude ($\sim$ 300 {\kms}) than for the typical ambient solar wind measured by MMS. Additionally, there is very little transverse motion of the plasma, $v_y$ and $v_z$ are $\sim$ 0. Similar to at other spacecraft, there is a density depletion (panel (c)) during the SA stream in comparison to the typical density seen while MMS is in the solar wind (red dashed line). Plasma beta ranges from $\sim$1 to $\sim$10 during this period, meaning that the plasma is dominated by kinetic pressure at MMS. $\mathrm{He^+/H^+}$ and $\mathrm{He^{++}/H^+}$ values are $\sim 10^{-3}$ and $\sim 10^{-1}$ respectively during the SA stream. There is a $\mathrm{He^{++}}$ enhancement on each side of the density depletion that should be further investigated.



\section{Turbulence Analysis} \label{sec: turbulence}

We calculate a variety of turbulence properties to study the dominant wave modes in the plasma during our period of interest. Specifically, we are interested in determining the {\alfty} level of the plasma, and the spectral properties of the {\elas} variables. We look for deviations in the dominant component in turbulence spectra from the ambient/supersonic solar wind which would imply heating by dissipation of low frequency turbulence. 


\subsection{Turbulence Parameters} \label{sec: turb-params}

We calculate two main normalized turbulence quantities: the cross helicity and residual energy. Cross helicity ({\sigmac}) is a measure of the degree of correlation between the fluctuations in the velocity and magnetic field and defined by Equation~\ref{eqn: sigmac}, where $E_\pm$ is the energy associated with the {\elas} variables, defined below. This can be used as a proxy for the {\alfty} of the plasma -- higher absolute cross helicity means the plasma is more {\alfic} (larger degree of correlation between the magnetic and velocity fields).
\begin{equation}\label{eqn: sigmac}
    \sigma_C = \frac{E_+ - E_-}{E_+ + E_-}
\end{equation} 

The normalized residual energy ({\sigmar}) is a measure of the dominant energy mode in the plasma and is defined by Equation~\ref{eqn: sigmar}. We expect $\sigma_C^2 + \sigma_R^2 \leq 1$ \citep{Bavassano-1998}, and therefore high {\alfty} plasma is balanced in energy.

\begin{equation}\label{eqn: sigmar}
    \sigma_R = \frac{\langle \delta \mathbf{v}^2 \rangle - \langle \delta \mathbf{b}^2 \rangle}{\langle \delta \mathbf{v}^2 \rangle + \langle \delta \mathbf{b}^2 \rangle}
\end{equation} 

The {\dv} and {\db} values are a measure of the fluctuations in the velocity and magnetic field respectively. They are calculated by subtracting the velocity (magnetic) field measurement from the mean background velocity (magnetic) fields which are determined by averaging over a 60-minute time interval to fully capture the spectral features in the inertial range. {\dv} is the fluctuation in the measured proton velocity vector (RTN coordinates) while {\db} is the fluctuation in the measured RTN magnetic field in {\alf} units: $\delta \mathbf{b} = \delta \mathbf{B}/\sqrt{\mu_0 \langle N_p \rangle m_p}$ using density observations from SPAN-I averaged over a 10-minute moving window and PSP/FIELDS magnetic field measurements. We time average ($\langle \cdot \cdot \cdot \rangle$) over non-overlapping 60-minute windows,chosen based the correlation times of Parker magnetic field fluctuations \citep{Parashar-2020, Chen-2020}. We also calculate the forward and backward-propagating {\elas} variables which are defined as: $\mathbf{z}^\pm = \delta \mathbf{v} \mp \mathrm{sign} \langle B_{r} \rangle \delta \mathbf{b}$ where $\mathrm{sign} \langle B_{r} \rangle$ is the averaged polarity of the magnetic field.

\begin{figure} [htb!]
  \includegraphics[width=\columnwidth]{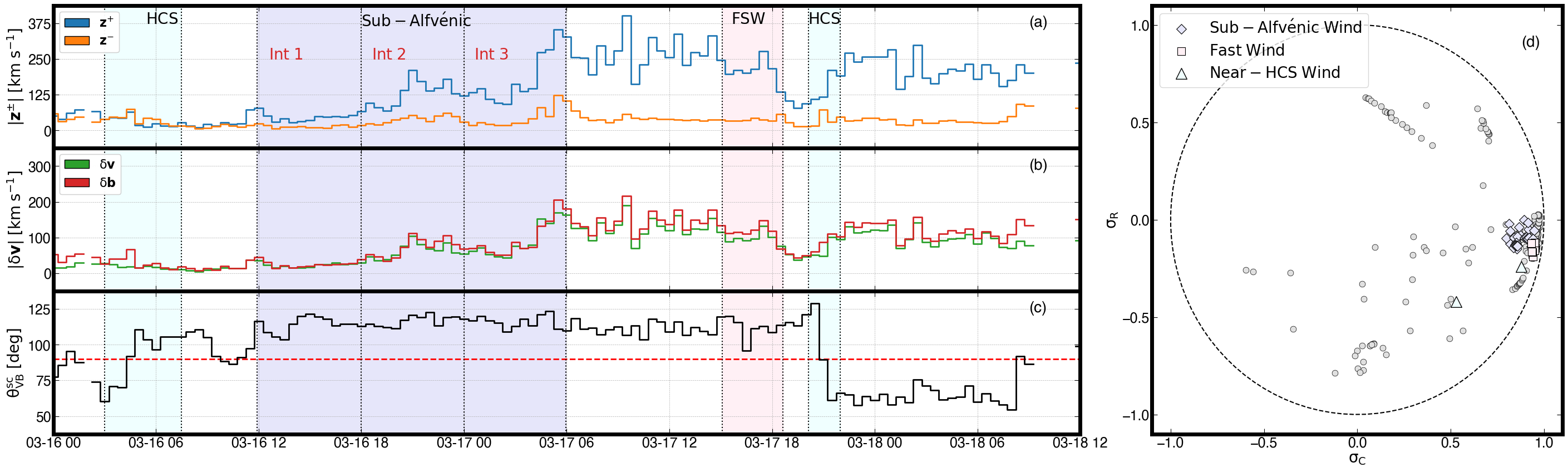}
  \caption{
    Turbulence parameters for Parker E15 computed over 20 minute non-overlapping windows following the description in Section~\ref{sec: turb-spectra} from SWEAP/SPAN-I and FIELDS measurements. The SA region is identified in purple, FSW region in pink, and the HCS crossings are shown in light blue. The two intervals (during the SA period) used for the PSD calculation in Figure~\ref{fig: turbulence_psd} are labeled as \lq{}Interval One\rq{} and \lq{}Interval Two\rq{}.
  \textit{Panel (a):} Elasser variables ({\kms}) showing the forward ({\zp}) and backward ({\zm}) modes in blue and orange respectively.
  \textit{Panel (b):} Velocity ({\dv}) and magnetic field ({\db}) fluctuations calculated by subtracting the measured velocity and magnetic fields from the background field.
  \textit{Panel (c):} The angle between the magnetic field and velocity in the spacecraft frame ({\angvb}).
  \textit{Panel (d):} Cross helicity ({\sigmac}) versus residual energy ({\sigmar}) for this encounter. FSW is shown in pink squares, SA wind in purple diamonds, and near-HCS wind in blue triangles. All other wind periods are in grey circles. The dashed circle gives the outline for our expected result that {\sigmac} + {\sigmar} $\leq$ 1. 
  }
  \label{fig: turbulence}
\end{figure}

Figure~\ref{fig: turbulence} shows an overview of the {\elas} variables, velocity ({\dv}) and magnetic field ({\db}) fluctuations, along with the cross helicity and residual energy. We split the SA period into three six-hours intervals to further study their turbulent properties. The intervals are labeled in Figure~\ref{fig: turbulence} and span from 2023-03-16/12:00 to 18:00, 2023-03-16/18:00 to 2023-03-17/00:00, and 2023-03-17/00:00 to 06:00.

During E15, there is large variation in the cross helicity and residual energy. When the absolute cross helicity $|\mathrm{\sigma_C}|\sim 1$, we expect pure {\alf} fluctuations near the Sun and $\mathrm{|\sigma_R| \sim 0}$, meaning the turbulence is balanced between the kinetic and magnetic energy \citep{Kasper-2019}. Conversely, {\sigmar}$\sim \pm 1$ implies the plasma is dominated by either kinetic (or magnetic) fluctuations for positive (negative) values of {\sigmar}. As expected by \citet{Shi-2022}, we see that the near-HCS plasma shows small {\sigmac} with negative {\sigmar} meaning the plasma is non-{\alfic} and dominated by magnetic fluctuations (panel (d)). 


Within the SA interval, we see negative {\sigmac} values meaning inward propagating {\alf} waves dominate this plasma.  During SA interval one, the inward propagating wave mode is almost entirely absent, and is unlikely to be a physical measurement (see Appendix~\ref{sec: appendix-turb}). The values for the Elasser variables during the SA period are quite low relative to the rest of the period. The velocity and magnetic field fluctuations ({\dv} and {\db}) are similar, and relatively low for this entire period, with slight dominance by the {\db} fluctuations. The residual energy during this period ranges between $\pm$0.25, but is primarily negative implying slight dominance by the magnetic energy. Similar to \citet{Zhao-2022}, we see that the magnetic pressure dominates in this very low $\beta$ region and as such we would expect negative {\sigmar}. Similarly, the FSW region is highly {\alfic} and dominated by magnetic energy fluctuations (negative {\sigmar}).


\subsection{Turbulence Spectra} \label{sec: turb-spectra}

\begin{figure} [h!]
  \includegraphics[width=\columnwidth]{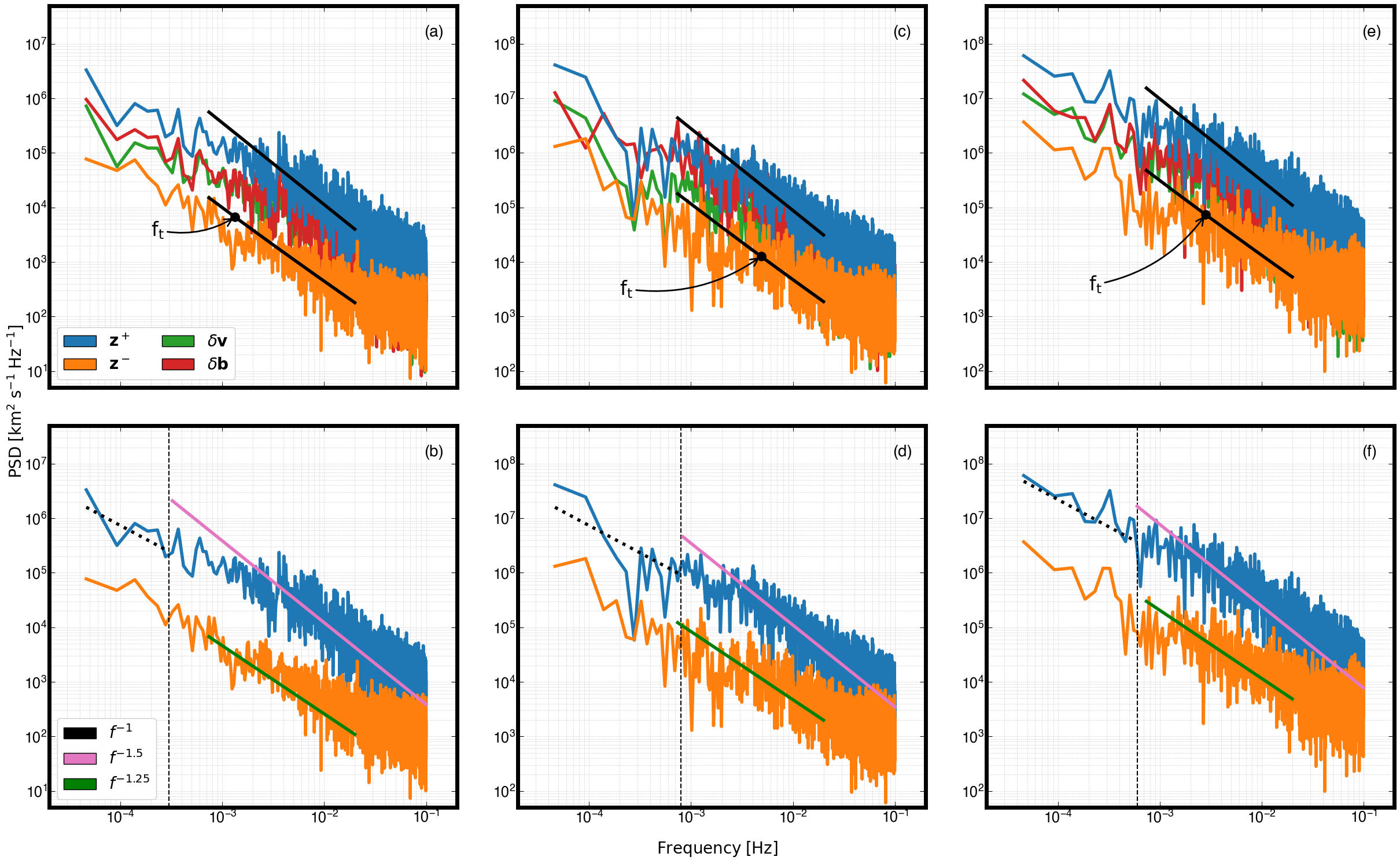}
  \caption{PSD of the turbulence parameters shown in Figure~\ref{fig: turbulence} for the SA period of interest. We split the period into three intervals (as seen in Figure~\ref{fig: turbulence}) and the PSD for each interval are shown in the left, middle, and right columns respectively.
  \textit{Top row:} Trace power spectra of the {\elas} variables ($\mathrm{\mathbf{z}^{\pm}}$) and fluctuations ({\dv} and {\db}) for interval one (panel (a)), interval two (panel (c)), and interval three (panel (e)). Fitted spectra for both {\zp} and {\zm} are shown in black \citep{Zank-2017, Zank-2022, Zhao-2022}. $f_t$ is the transition frequency from nonlinear to {\alfic} interaction dominated regions.
  \textit{Bottom row:} PSD of the inward ({\zm}) and outward ({\zp}) propagating {\elas} variables showing spectral fits. The dashed line separates the two regimes ($f^{-1}$ and $f^{-1.5}$) of the {\zp} spectra, where the $f^{-1}$ portion at small frequencies corresponds to the energy containing range. This is related to the correlation time scale of {\zp} during each period \citep{Zhao-2022-corr}.
  }
  \label{fig: turbulence_psd}
\end{figure}

In Figure~\ref{fig: turbulence_psd}, we show the power spectral density (PSD) for the {\elas} variables along with for {\dv} and {\db} during our SA interval. We calculate the trace power spectra by applying a Tukey window ($\alpha$=0.5) \citep{Harris-1978} and then a Fourier transform to each turbulence parameter. We divide our 18-hour period into three 6-hour periods and calculate the PSD over each interval. We show the spectra for each interval separately due to the differences in amplitudes of the turbulence parameters in each interval. The PSD is normalized following Equation~\ref{eqn: psd} where $f_s$ is the sampling frequency, $S[n]$ is the power spectra, and $\tilde{x}[k]$ is the Fourier transform of the timeseries $x[n]$ \citep{Paschmann-1998}.

\begin{equation}\label{eqn: psd}
    S[n] = \frac{1}{W_{ss}}\frac{2N}{f_s}[\tilde{x}[k]]^2
\end{equation}

In order to compare with the methods of \citet{Zank-2020, Zank-2022, Zhao-2022}, we fit our observed PSD of the {\elas} variables with a nearly incompressible (NI) turbulence model \citep{Zank-2017} which includes both a quasi-2D MHD component ($f^{-5/3}$ power law) and a slab component ($f^{-3/2}$ power law). The quasi-2D MHD component is a non-propagating (advecting) component following the classic $f^{-5/3}$ power law \citep{Kolmogorov-1941}. It is assumed to contain equal power in both the {\zp} and {\zm} modes and as it is non-propagating, there is no Doppler shift between inward and outward modes \citep{Zank-2022, Zhao-2022}. The slab component describes the inward and outward propagating {\alfic} fluctuations aligned with the mean magnetic field \citep{Zank-2018}, and includes a Doppler shift of the {\zp} and {\zm} modes: $k_+/k_- = (|u_0 \cos{\Phi} + V_{A0}|)/(|u_0 \cos{\Phi} - V_{A0}|)$ where $u_0$ is the mean speed in the spacecraft frame and $V_{A0}$ is the mean {\alf} speed during the SA period. In a field aligned case ({\angvb}$\sim 0$), the slab component is dominant \citep{Zhao-2022}, but for a highly oblique angle between the mean magnetic field and mean solar wind speed ({\angvb}) both the slab and quasi-2D components are important. In Figure~\ref{fig: turbulence}, we show the highly oblique nature of this stream and thus we can compare the contribution of the components following the form of the NI MHD turbulence model as in \citet{Zhao-2022} to describe the {\elas} spectra:

\begin{center}
\begin{align} \label{eqn: spectral-fit}
    G_+(k+) = C^{* +}k_+^{-3/2} + C^{\infty}k_+^{-5/3} \\
    G_-(k-) = C^{* -}k_-^{-3/2}\left(1 + \sqrt{\frac{k_-}{k_t}} \right)^{1/2} + C^{\infty}k_-^{-5/3}. \label{eqn: spectral-fit-minus}
\end{align}  
\end{center}
$k_{\pm} = 2 \pi f / (|u_0 \cos{\Phi} \mp V_{A0}|)$, $C^{* \pm}$ gives the power in the slab mode, $C^{\infty}$ gives the power in the quasi-2D fluctuations and $\pm$ refers to the inward ({\zm}) and outward ({\zp}) propagating {\elas} variables.

Figure~\ref{fig: turbulence_psd}, shows the power spectra of the {\elas} variables ({\zp} and {\zm}), and fluctuations in the velocity ({\dv}) and magnetic ({\db}) fields, which are calculated following the methodology described in Section~\ref{sec: turb-spectra}. We calculate the PSD for the three identified intervals separately, as these three regions show different fluctuation amplitudes.  The intervals for the PSD calculation are 2023-03-16/12:00 to 18:00, 2023-03-16/18:00 to 2023-03-17/00:00, and 2023-03-17/00:00 to 06:00, as shown in Figure~\ref{fig: turbulence}.

We fit the {\elas} variable spectra using the methods described above and find the following description of the spectra (Table~\ref{tab: spectral-fit}). These coefficients are determined by fitting Equations~\ref{eqn: spectral-fit} and \ref{eqn: spectral-fit-minus} where $C^{* \pm}$ and $C^{\infty}$ describe the power in the slab mode and in the 2D MHD mode respectively. The transition wave number (frequency) $k_t$ ($f_t$) is where the spectra flattens from a steeper slope ($f^{-5/3}$) controlled by non-linear (quasi-2D) interactions when $k > k_t$ to being dominated by {\alfic} wave-like (slab turbulence) interactions ($f^{-3/2}$) for $k < k_t$. Together, these coefficients describe the fit shown by the black lines in Figure~\ref{fig: turbulence_psd} based on Equations~\ref{eqn: spectral-fit} and \ref{eqn: spectral-fit-minus} . 

\begin{table}
  \centering
  \begin{tabular}{|c||c|c|c|c|c|}
    \hline
    \textbf{Interval} & \textbf{Time Range} & \textbf{$C^{* +}$} & \textbf{$ C^{* -}$} & \textbf{$C^{\infty}$} & \textbf{$f_t$} [Hz] \\
    \hline
    \textbf{Interval One} 
        & 2023-03-16/12:00 to 18:00
        & $\mathrm{1.6 \times 10^{-2}}$ 
        & $\mathrm{7.7 \times 10^{-4}}$
        & $\mathrm{4.7 \times 10^{-14}}$
        & $\mathrm{1.3 \times 10^{-3}}$\\
    \hline
    \textbf{Interval Two}
        & 2023-03-16/18:00 to 2023-03-17/00:00
        & $\mathrm{2.8 \times 10^{-2}}$ 
        & $\mathrm{1.5 \times 10^{-3}}$
        & $\mathrm{1.0 \times 10^{-15}}$ 
        & $\mathrm{4.9 \times 10^{-3}}$\\
    \hline
    \textbf{Interval Three} 
        & 2023-03-17/00:00 to 06:00
        & $\mathrm{1.4 \times 10^{-1}}$ 
        & $\mathrm{8.5 \times 10^{-3}}$
        & $\mathrm{7.1 \times 10^{-14}}$
        & $\mathrm{2.8 \times 10^{-3}}$\\
    \hline
  \end{tabular}
  \caption{Fit parameters for the NI turbulence model \citep{Zank-2020, Zank-2022} based on Equations~\ref{eqn: spectral-fit} and 5. $C^{* \pm}$ corresponds to the power in the slab component ($f^{-3/2}$)  for the inward ({\zm}) and outward ({\zp}) propagating modes. $C^{\infty}$ is the power in the quasi-2D fluctuations ($f^{-5/3}$). The transition frequency $f_t$ (wave number $k_t$) separates the strong (quasi-2D interactions) and weak ({\alf} wave propagation dominated) turbulence regions.}
  \label{tab: spectral-fit}
\end{table}

We look at the ratio of the coefficients in Table~\ref{tab: spectral-fit} to determine the dominant mode. For the outward ({\zp}) mode we find $C^{\infty}_i / C^{* +}_i  <<< 1$ for all intervals (i = 1, 2, 3). The same is seen for the inward mode ({\zm}), where $C^{\infty}_i / C^{* -}_i  <<< 1$, meaning that both the inward and outward modes are dominated by the slab ($f^{-3/2}$) component in this model. The outward propagating {\zp} fluctuations dominate with a spectral amplitude nearly an order of magnitude larger than the {\zm} mode, meaning that this SA period is dominated by outwardly propagating {\alf} waves. While the scaling we find for these spectra are compatible with this model, they are also compatible with other phenomenological turbulence theories based on the principles of \lq{}critical balance\rq{} \citep{Goldreich-1995, Goldreich-1997} and \lq{}scale-dependent dynamic alignment\rq{} \citep{Boldyrev-2006, Chandran-2015, Mallet-2017}.


In the bottom row, we show overall fits to the {\zp} and {\zm} spectra. We note that the {\zp} shows two regimes: a $\sim f^{-1}$ dependence at lower frequencies (in the energy containing range), then a spectral break (black dashed line) to a steeper $\sim f^{-3/2}$ power law. This break (shown by the dashed line in Figure~\ref{fig: turbulence_psd}) is the frequency corresponding to the correlation scale separating the inertial and energy containing range \citep{Zank-2022}. This is similar to results seen in \citet{Zank-2022} for SA wind. The $f^{-1}$ fit in the low frequencies of the {\zp} spectra has been observed by \citet{Zank-2022} and possible explanations for this include: saturation of the fluctuations amplitude at larger scales and that this break between the $f^{-1}$ and $f^{-1.5}$ occurs when $<\delta B / B \sim 1>$ \citep{Matteini-2018}, an inverse energy cascade driven by \lq{}anomalous coherence\rq{} effects \citep{Velli-1989, Meyrand-2023} and/or the  parametric decay instability \citep{Chandran-2018}.

The {\zm} spectra is fit over the inertial range ($\mathrm{5 \times 10^{-4}}$ to $\mathrm{1 \times 10^{-2}}$ Hz) and shows behavior distinct from the the {\zp} mode. These spectra show a spectral index decrease from a $f^{-3/2}$ spectra at smaller scales to a $f^{-1.25}$ dependence until flattening around $10^{-2}$ Hz. This $f^{-1.25}$ dependence at larger scales is to be expected from Equation~\ref{eqn: spectral-fit-minus}. As shown in Table~\ref{tab: spectral-fit}, the {\zm} mode is dominated by the $f^{-3/2}$ component, meaning that at larger scales, this spectra can be approximated as $G_-(k-) \propto k_-^{-1.25}$. Recent work by \citet{Wu-2024}, has shown flattening of the spectral slope in the {\zm} spectra near the Sun. As we describe below and in Appendix~\ref{sec: appendix-turb}, it is important to look at the sources of noise in the {\zm} spectra to determine whether fitted spectral slopes and flattening of the spectra at higher frequencies are physical.


There are a variety of sources of noise in the calculation of the fluctuations and {\elas} variables that can lead to uncertainty in these measurements and flattening of the spectra. For example, noise in the density measurements used to scale {\db}, noise introduced by data compression,  digitization, and Poisson statistics in the SPAN-I data processing pipeline, and the finite velocity grid of the SPAN-I measurements. In Appendix~\ref{sec: appendix-turb}, we outline our quantification of the effect of the finite velocity grid of the PSP/SPAN-I measurements to the relation between {\zm} and {\dv} to discuss whether the observed spectral flattening of the {\zm} and {\dv} spectra at $\sim 10^{-2}$ Hz is physical. If this were a physically measurable signal, it would imply that these streams have more energy at smaller (higher) scales (frequencies). Therefore, determining whether this is a real or instrumental signal is important to understand whether we can infer physical mechanisms from the flattening of these spectra. We take the SPAN-I energy bins in velocity space and use the mean observed velocity during the interval to estimate the minimum measurable fluctuation during the specified periods. From this analysis, we determine that the flattening of the {\zm} spectra is due to this finite velocity grid, along with other sources of noise not discussed in this paper, rather than a physical effect. The quantification of the noise floor and comparison with the observed spectral flattening means that we cannot make conclusions about the spectral shape or dominant wave modes at higher frequencies for this SA stream. It is important to remember that there are a variety of sources of noise or error in the measurements of the {\zp} and {\zm} modes, and any conclusions based on these spectra should discuss these effects and their impact on the conclusions made.

\section{Source Region Connection} \label{sec: source-region}

In Section~\ref{sec: data}, we outlined the in situ observations from Parker during our period of interest. We will now compare these measurements with the results of the modeling and calculations described below. Connecting observations to theoretical results is a primary science goal of the Parker Solar Probe and {\SO} missions, and shows the power of multi-point observations to provide invaluable insight to our understanding of the solar wind. Due to the potential connection of these low density, sub-{\alfic} intervals to transient CME structures, we discuss basic modeling of the propagation of a CME that erupted prior to this SA stream and its relation to the observations.

\subsection{Relation to Transients} \label{sec: disc-transients}

There is ongoing work looking at the connection of the low density, SA streams to transient structures such as the rarefaction region or post-flow depletion region of a CME. There was a small CME of March 15 06:48 which erupted from an AR very close to the AR source of the SA stream. In order to determine if there is a relation between this transient structure and the SA stream, we use an ice cream cone model\citep{Na-2017} to propagate the CME from its eruption out to 1AU. We compare the propagation of the CME with observations of the low density stream at 1AU shown in Figure~\ref{fig: radial} to determine if there is a connection.

We identify the source region of the March 15 CME through AIA and LASCO coronagraph images and assume the CME propagates radially outward from the AR in the inertial coordinate system.  Using coronagraph observations, we determine the angular width and propagation speed of the CME, then use this to evolve the eruption through time based on the simplified ice cream cone model in an inertial coordinate frame \citep{Na-2017}. We compare this with the trajectories of Parker, {\SO}, and 1AU spacecraft in the inertial frame. In the inertial coordinate frame, solar wind streams rotate through time due to solar rotation. Thus the SA stream rotates around the Sun due to solar rotation in the inertial frame, but is stationary in the Heliographic Carrington frame as in Figure~\ref{fig: radial}. A full resolution video of this simulation can be found at \citet{Ervin-2024-CME}.


\begin{figure} [ht] 
\begin{center}
  \includegraphics[width=0.6\textwidth]{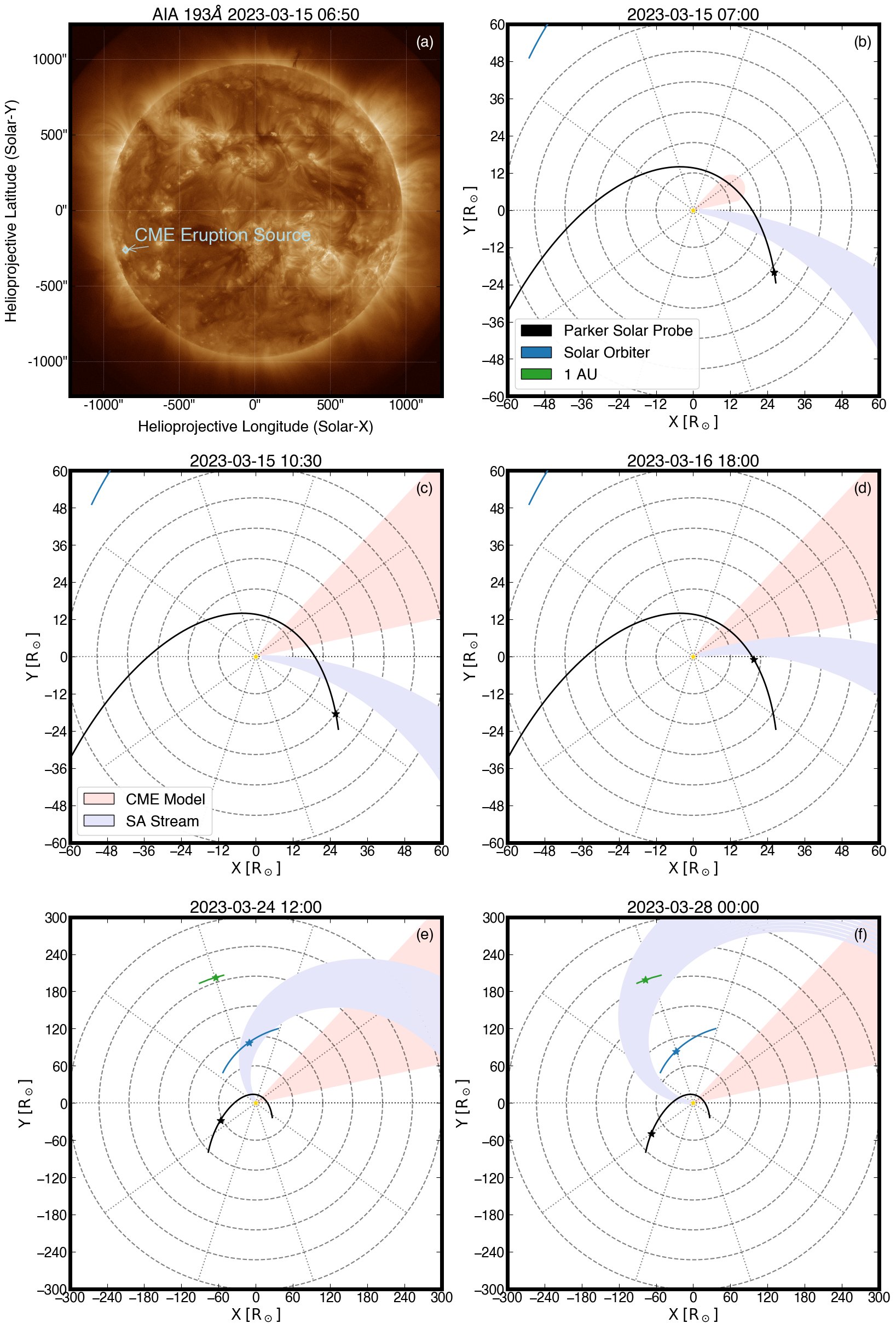}
  \caption{Comparison of spacecraft trajectory and the SA stream with CME propagation based on a simplistic ice cream cone model \citep{Na-2017} in the inertial coordinate frame. Each panel shown is a still from an animation showing the evolution of the SA stream and CME during the entire Encounter 15 period \citep{Ervin-2024-CME}. The animation spans from March 15 to April 3, 2024 lasting 3:02 and is annotated identically to the panels in this Figure with the addition of timestamps along each spacecraft trajectory. The CME trajectory is shown in pink, the SA stream is in purple, and spacecraft trajectories are in black (Parker), blue (Solar Orbiter), and green (1AU). The stars in each panel indicate the location of the spacecraft at each time step.
  \textit{Panel (a):} AIA 193{\AA} image from March 15 06:48 showing the AR source of the CME eruption. 
  \textit{Panel (b):} March 15 07:00 showing the onset of the CME eruption.
  \textit{Panel (c):} March 15 10:30 showing the extent of the CME eruption.
  \textit{Panel (d):} March 16 18:00 showing Parker flying through the SA stream. 
  \textit{Panel (e):} March 24 12:00 showing {\SO} flying through the SA stream.
  \textit{Panel (f):} March 28 00:00 showing 1AU spacecraft flying through the SA stream.
  }
  \label{fig: cme}
\end{center}
\end{figure}

In Figure~\ref{fig: cme}, we show five different time steps of our CME model and SA stream during the period of interest. Panel (a) shows the location of the CME eruption on an AIA 193{\AA} observation based on the LASCO catalog. This is the point we use at the source for our ice cream model. The follow time steps (panel (b) and (c)), show the onset and extent of the CME eruption. This eruption passes through the Parker trajectory prior to Parker encountering the wake region of the CME and thus Parker should not have observed this CME in situ. The CME also does not intersect with the Solar Orbiter or 1AU spacecraft trajectories. In panel (d), Parker is shown midway through the spacecraft observations of the stream corresponding to the low {\MA} and large density depletion. In Panel (e), the stream intersects with the {\SO} trajectory however there are no in situ observations from {\SO} to compare with this. Panel (f) shows the intersection with the 1AU trajectories (Wind, MMS) at the time periods of the large density dip and near $\sim 1$ {\MA} observed. While this CME propagation model is quite simplistic, it captures the overall heliospheric structure surrounding this stream and shows that these multi-spacecraft observations are likely due to a single source rather than interaction with transients. 


\subsection{Magnetic Field Modeling} \label{sec: methods-model}
We use global coronal models to connect our in situ observations with a coronal source region. When used in combination the in situ particle, magnetic field, and composition metrics, this allows for a characterization of the source region of the plasma. Using PFSS and MHD models, we can create photospheric footpoint estimates of where the plasma that Parker and other spacecraft measure during this time period originates from. Using multiple modeling methods allows for cross comparison both between the models themselves, and with in situ constraints, leading to confidence in footpoint estimations.

The PFSS modeling method is a computationally efficient method for a creating a global model of the coronal magnetic field that takes an observed photospheric radial magnetic field as an input \citep{Schatten-1969, Altschuler-1969} and assumes a current-free or magnetostatic ($\nabla \times \mathbf{B} = 0$) corona. The assumption of a magnetostatic corona coupled with Maxwell's equation stating $\nabla \cdot \mathbf{B} = 0$ means the magnetic field can be solved as a scalar potential: $\mathbf{B} = - \nabla \Phi_B$ using the Laplace equation ($\nabla^2 \Phi_B = 0$) out to an outer boundary. The outer boundary is an equipotential surface (known as the radial source surface -- {\Rss}) where magnetic field lines are forced to be purely radial, thus modeling the effect of the magnetic field on the outflow of the solar wind. 

We use the \texttt{pfsspy} package \citep{pfss} to create PFSS models using synoptic ADAPT-GONG magnetograms as the input boundary condition. \texttt{pfsspy} creates a full 3D magnetic field model and can trace magnetic field lines from a uniform photospheric grid out to the {\Rss} height. We compare PFSS results with Parker and {\SO} radial magnetic field measurements to determine the best model instantiation via comparison of the modeled HCS with the observed neutral line crossings, determining 2.5{\Rsun} as the optimal source surface height for this period. In order to connect the spacecraft locations (between 13.3{\Rsun} and 120{\Rsun}) to the source surface, we use ballistic propagation to map fields line from the spacecraft trajectory to the {\Rss} and connect with the model  \citep{Snyder-1966, Nolte-1973, Macneil-2022, Koukras-2022, Badman-2020}. 

The combination of ballistic propagation and a global, coronal magnetic field model allows us to estimate the footpoints from which the observed plasma emerged. To understand how errors in the PFSS and ballistic propagation models affect our results, Appendix~\ref{sec: appendix-model-val} compares the estimated footpoints at varying source surface heights and with induced errors in ballistic propagation, and shows that the resulting source region does not change for our periods of interest. These footpoint estimations allow us to determine properties of the source region of the plasma, most notably the photospheric magnetic field value and footpoint brightness. Together with other observables, these parameters provide evidence of the source region of the solar wind. 


In addition to the simple potential approximation of the coronal magnetic field implemented by the PFSS model, we use an MHD modeling approach to understand the structure of the global coronal field. We use the Predictive Science Inc. (PSI) Magnetohydrodynamic Algorithm outside a Sphere (MAS) code which integrates the time-dependent thermodynamic MHD equations over a 3D non-uniform mesh grid \citep{Riley-2021}{\mascode}. MAS requires photospheric radial magnetic field observations and a heating mechanism as input conditions to create a realistic 3D description of the solar corona. The MAS model used in this study implements empirical thermodynamic approximations of energy transport to solve the MHD equations \citep{Lionello-2001, Lionello-2009} in two regimes: the coronal regime (1{\Rsun} to 30{\Rsun}) and a heliospheric regime (30{\Rsun} to 1AU). The resulting 3D solution can be sampled along the spacecraft's trajectory to compare the structure of synthetic observables produced by the model (magnetic field, density, velocity) to in situ observations.

To verify the validity of our models, we compare HCS estimations between the PFSS and MHD modeling methods and compare modeled observables with in situ measurements. From panel (a) in Figure~\ref{fig: ar-detection}, we see the modeled HCS from the PFSS solution shows strong alignment with the observed polarity reversal measured by Parker (change in color of the Parker trajectory and panel (b) results). Additionally, in Appendix~\ref{sec: appendix-model-val}, we show the same alignment between the modeled HCS from the MHD and PFSS solutions with the Parker and {\SO} measured polarity reversals. We show a direct comparison of the observables produced by the MHD model (proton radial velocity, radial magnetic field, and proton density) with Parker measurements finding that the synthesized observables show similar structure to the observations (Figure~\ref{fig: mhd_obs}). These validation methods of comparing models to in situ constraints and comparing the models with each other, provide strong confidence in our photospheric footpoint estimates and subsequent discussion of solar wind source regions.



In addition to estimating footpoints and the structure of the global coronal field, the models allow for the calculation of the magnetic expansion factor which has been tied to solar wind source region. The magnetic expansion factor ($\mathrm{f_s}$) was first defined by \citet{Wang-1990} and provides the basis for the Wang-Sheeley-Arge (WSA) solar wind prediction model \citep{Arge-2000, Arge-2003, Arge-2004} due to its empirically determined inverse correlation with solar wind speed \citep{Wang-1990, Riley-2012, Riley-2015}. It relates the expansion (solid angle) of a flux tube from one location (nominally the solar surface at 1.0{\Rsun}) to some other point in the corona $\mathrm{(R_1, \lambda_1, \phi_1)}$, and is defined as in \citet{Wang-1997} by Equation~\ref{eqn: fss}:

\begin{equation}\label{eqn: fss}
    f_{s} = \left( \frac{R_0}{R_1}\right)^2 \frac{B_R(R_0, \lambda_0, \phi_0)}{B_R(R_1, \lambda_1, \phi_1)}
\end{equation} 

In our work, we calculate the magnetic expansion factor from the photosphere (1.0{\Rsun}) to the source surface (2.5{\Rsun}) where field lines are assumed to be fully radial. For a flux tube that passes through a point on the source surface $\mathrm{(R_1, \lambda_1, \phi_1) = (R_{ss}, \lambda_{ss}, \phi_{ss})}$, we define the expansion factor as {\fss} which is a measure of the expansion of the flux tube between its photospheric footpoint and the source surface. When {\fss} = 1, the open field lines diverge as $r^2$, and values greater than 1 imply more rapid divergence \citep{Wang-1997}. 


\subsection{Source Region Detection} \label{sec: disc-detection}
\begin{figure}[ht]
\begin{center}
  \includegraphics[width=0.8\columnwidth]{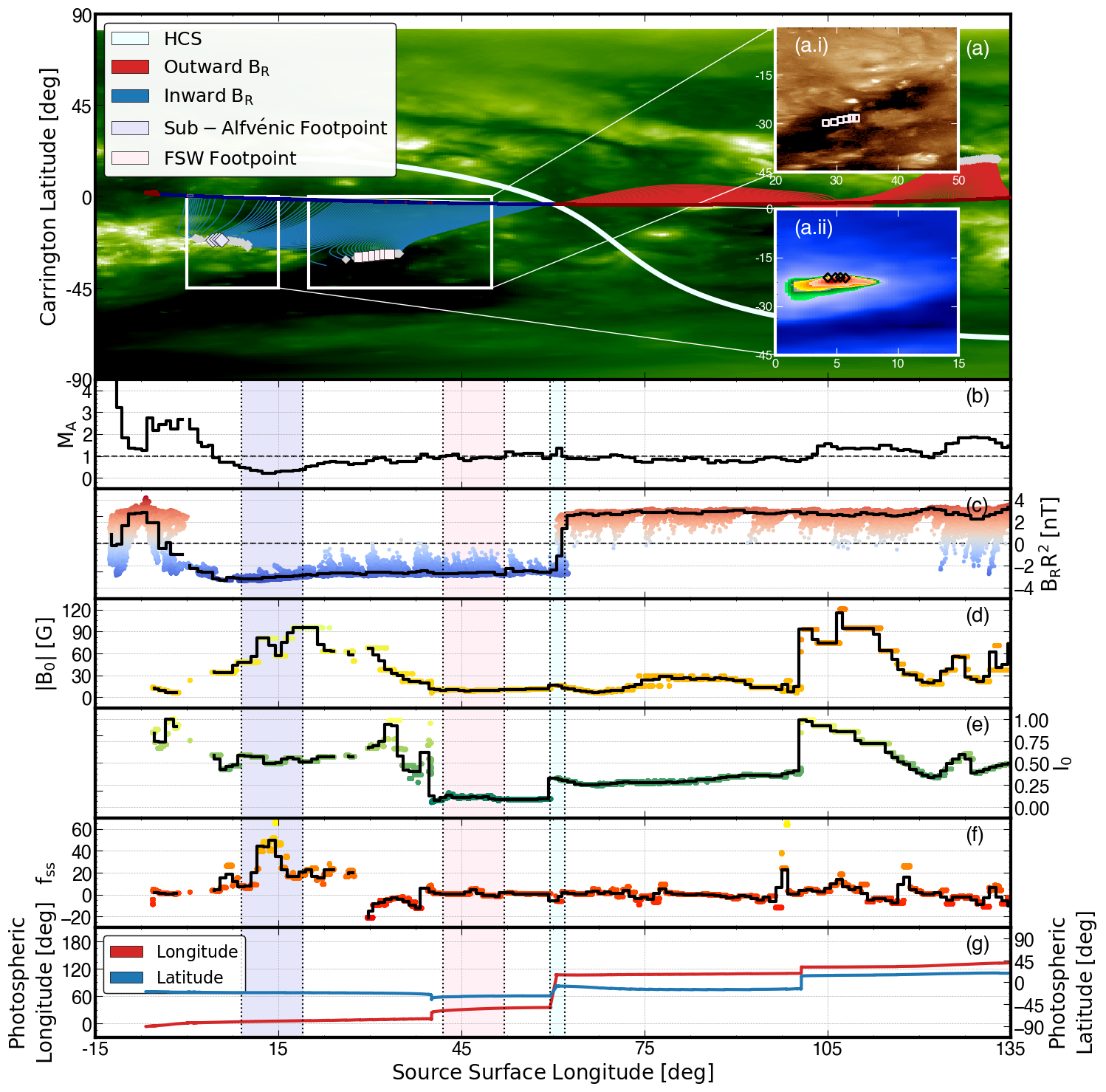}
  \caption{
    Modeling results compared with in situ measurements and calculations with ballistic propagation to show data as a function of source surface longitude. The raw data is overlaid with data binned by 1{\degree} in source surface longitude (black).
    \textit{Panel (a):} PFSS model showing the modeled HCS (white) and estimated photospheric footpoints. We ballistically propagate the Parker trajectory, colored by measured polarity, to the source surface (2.5~{\Rsun}) using a varying solar wind speed (the SPAN-I $\mathrm{v_R}$). The modeled field lines, colored by modeled polarity, connect the spacecraft trajectory to the photospheric footpoints. We overlay this model on a Carrington map produced from SDO/AIA 193{\AA} images. 
    \textit{Panel (a.i):} FSW footpoints as pink squares overlaid on a SDO/AIA image from the estimated time when the plasma left the Sun based on the measured wind velocity and location of the spacecraft.
    \textit{Panel (a.ii):} GONG photospheric radial magnetic field with the SA footpoints overlaid as black diamonds.
    \textit{Panel (b):} {\alf} Mach number ({\MA}) as a function of source surface longitude in 1{\degree} bins. The dashed line at 1 shows where wind becomes sub-{\alfic}.
    \textit{Panel (c):} The scaled radial magnetic field as measured by Parker/FIELDS. The dashed line at 0 shows the crossing of the neutral line by the spacecraft.
    \textit{Panel (d):} The absolute photospheric radial magnetic field ($\mathrm{|B_{r, 0}|}$) based on estimated footpoints in panel (a). 
    \textit{Panel (e):} The relative intensity of the photospheric footpoints ($\mathrm{I_0}$) based on estimated footpoints in panel (a). This can be thought of as a proxy for the temperature of the plasma at the footpoint (darker/lower is colder).
    \textit{Panel (f):} Magnetic expansion factor ($\mathrm{f_{ss}}$) between the source surface ({\Rss} = 2.5~{\Rsun}) and photosphere (1.0~{\Rsun}) calculated following Equation~\ref{eqn: fss}. 
    \textit{Panel (g):} Photospheric footpoints in longitude (red) and latitude (blue) for Parker calculated using a PFSS model. Jumps in footpoints are indicative of a change in the structure Parker is magnetically connected to.
  }
  \label{fig: ar-detection}
\end{center}
\end{figure}

In addition to the in situ particle, magnetic field, and composition metrics discussed previously, modeling results show that this SA, sub-sonic, slow solar wind stream emerges from a highly magnetized active region. In Figure~\ref{fig: ar-detection}, we compare modeling results with parameters derived from our models, described in Section~\ref{sec: methods-model}, as a function of source surface longitude. Like in Figure~\ref{fig: radial}, we use ballistic propagation to map the spacecraft location from its trajectory to the {\Rss}. In addition to the SA interval, we look at a FSW interval (highlighted in pink) to ensure that the resulting source region matches with where we expect FSW to originate -- CHs \citep{vonSteiger-2000, McComas-1998, McComas-2008}.

We study the fast wind region from 42{\degree} to 52{\degree} (in source surface longitude) as a means to validate our modeled footpoint estimations, expansion factor calculation, and as comparison with in situ observations. The wind during this time period has speeds of $\sim$400 {\kms} from Parker and $\sim$700 {\kms} at Solar Orbiter. The estimated footpoints are shown in pink squares in Figure~\ref{fig: ar-detection}, filling a CH region. The relative footpoint brightness (EUV intensity) and absolute magnitude of the photospheric radial field is low, while the magnetic expansion factor is $\sim$1. Low expansion factor is typical for CH wind due to the inverse relationship between wind speed and {\fss} \citep{Wang-2009}. The low relative footpoint brightness in EUV images (darker pixel) means cooler, lower density plasma which is indicative of CHs \citep{Cranmer-2009}. CHs are also the least active regions on the Sun \citep{Cranmer-2009}, as seen by the incredibly low photospheric field strength in this period. In panel (e), there is a jump in source surface footpoints before and after this region, indicative of CH boundaries. This analysis showing the FSW stream emerging from a CH region, coupled with matching the neutral line crossings gives us confidence in our picture of the overall global coronal field \citep{vonSteiger-2000, McComas-1998, McComas-2008}. 

The SA region is identified in Figure~\ref{fig: ar-detection} between 9{\degree} and 19{\degree} (in source surface longitude), and in panel (a) the estimated photospheric footpoints are in purple mapping to strong photospheric field concentrations. The photospheric footpoints for the AR lie on top of an AR seen in EUV, at $\sim$ -20{\degree} in latitude, well within the $\pm$50{\degree} latitude range for previously observed ARs \citep{Stansby-2021aa}. In panel (c), the scaled radial magnetic field ($\sim$-3nT) is the strongest field measured during E15 and the photospheric radial magnetic field ($\mathrm{B_0}$) during this period reaches $\sim$-95G. \citet{Stansby-2021aa} studied the contributions of active regions to the solar wind, and found a threshold of 30G ($\mathrm{3 \times 10^6}$nT) between active region and CH origins in GONG magnetograms. The relative footpoint brightness (panel (e)) during this region is the highest of E15. Active regions are areas of strong magnetic fields on the solar surface and manifest in EUV images as bright regions (high footpoint brightness) in the corona, meaning the high relative brightness indicates AR origins.

We find the active region we map to has a large expansion factor, meaning its open field lines have a complex topology. This is consistent with the Wang-Sheeley relationship between slow wind and high expansion factor, suggesting a steady accelerating flow from this region, unlike typical slow wind from pseudostreamers \citep{Wallace-2020}.


In panel (g) we show the source surface ({\Rss} = 2.5{\Rsun}) footpoints for Parker. We see that there is a jump in footpoints at the HCS crossing, another indicator that validates the models and characteristic of the streamer belt. The footpoints in the SA region remain constant over the entire time period indicating that Parker was magnetically connected to the same active region throughout this period of interest.

This combination of in situ parameters and modeling results clearly show that the slow, SA wind period emerges from a highly magnetized active region at mid-latitudes. We see high relative footpoint brightness, the strongest measured scaled radial magnetic field and footpoint field strength of the period, and a high magnetic expansion factor indicative of active region origins.

\section{Results} \label{sec: results}
We combine multi-point measurements with modeling to characterize a near subsonic, extremely sub-{\alfic} solar wind stream and determine the properties that differentiate this stream from the ambient solar wind. This is the closest to subsonic solar wind that Parker has encountered and provides a laboratory to explore extreme solar wind conditions. We outline the plasma parameters for this SA slow wind stream noting that it is a highly incompressible stream dominated by magnetic pressure (low plasma $\beta$) and outward propagating {\alfic} fluctuations ({\sigmac} $\sim$ 1). It is important to note that the values of $\mathrm{\langle \sigma_C \rangle}$ shown in this table do not represent the true nature of the imbalance of the fluctuations due to the fact $\mathbf{\sigma_C}$ is calculated using single point measurements in the spacecraft frame, and because of the Doppler effect. We discuss the turbulent parameters in the context of determining the dominant wave modes in the SA stream as compared to typical modes observed in the solar wind. 

Using PFSS and MHD models, we trace the Parker observations to their estimated source region and find that this stream originated from a mid-latitude AR. We discuss a FSW stream and near-HCS plasma for comparison and find that the SA period shows dramatically different characteristics than these typical wind streams: turbulence dominated by slab component ($f^{-3/2}$) rather than the $f^{-5/3}$ power law, incredibly low alpha particle abundance, and a highly magnetized in situ magnetic field signature. We discuss the quantification of the SPAN-I noise floor and how this prevents us from inferring physical turbulence properties at high frequencies in these streams. 


Table~\ref{tab: parameters} outlines the in situ and modeled parameters for the FSW and SA periods of interest. This allows for direct comparison of the SA characteristics with more \lq{}typical\rq{} solar wind parameters. In this study of Parker E15, we find the following parameters for our regions of interest:

\begin{table}
  \centering
  \begin{tabular}{|c|c||c|c|c|c|}
    \hline
    \textbf{Parameter} & \textbf{Units} & \textbf{SA Region} & \textbf{FSW}  \\
    \hline
    \hline 
    Start Time & UT & 2023-03-16 11:55  &  2023-03-17 13:52 \\
    \hline
    End Time & UT & 2023-03-17 05:05 & 2023-03-17 18:37 \\
    \hline
    \textbf{$\mathrm{\langle M_A \rangle}$} & & 0.32 & 0.96 \\
    \hline
    \textbf{$\mathrm{\langle v_R \rangle}$} & {\kms} & 171.69 & 422.80 \\
    \hline
    \textbf{$\mathrm{\langle A_{He} \rangle}$} & & 0.005 & 0.047 \\
    \hline
    \textbf{$\mathrm{\langle v_{\alpha, p} \rangle}$} & & 0.09 & 0.42 \\
    \hline
    \textbf{$\mathrm{\langle B_R R^2 \rangle}$} & $\mathrm{nT \; AU^2}$ & -3.08 & -2.68 \\
    \hline
    \textbf{$\mathrm{\langle B_R(r=1R_{\odot}) \rangle}$} & G & -73.36 & -9.52 \\
    \hline
    \textbf{$\mathrm{\langle I_0 \rangle}$} & & 0.71 &  0.29 \\
    \hline
    \textbf{$\mathrm{\langle f_{ss} \rangle}$} & & 35.20 & 0.79\\
    \hline
    \textbf{$\mathrm{\langle \sigma_C \rangle}$} & & 0.79 & 0.89 \\
    \hline
    \textbf{$\mathrm{\langle \sigma_R \rangle}$} & & -0.04 & -0.19 \\
    \hline
    \textbf{$\mathrm{\langle \delta \mathbf{v}\rangle}$} & {\kms} & 31.5 & 97.8 \\
    \hline
    \textbf{$\mathrm{\langle \delta \mathbf{b}\rangle}$} & {\kms} & 34.3 & 119.6 \\
    \hline
    \textbf{$\mathrm{\langle \mathbf{z}^+ \rangle}$} & {\kms} & 64.7 & 215.39 \\
    \hline
    \textbf{$\mathrm{\langle \mathbf{z}^- \rangle}$} & {\kms} & 12.8 & 36.6 \\
    \hline
  \end{tabular}
  \caption{Observed and calculated parameters for the FSW SA and FSW periods. $\langle \cdot \cdot \cdot \rangle$ signifies an average of the parameter over the entire period. $\mathrm{I_0}$ is the relative photospheric footpoint brightness based on a synoptic map produced from SDO/AIA 193{\AA} disk images. SS Lon is the range in source surface longitude that the region of interest covers. $\mathrm{v_{\alpha, p}}$ is the normalized alpha-to-proton differential velocity.}
  \label{tab: parameters}
\end{table}

\begin{enumerate}

    \item We identify an 18-hour sub-{\alfic} and near subsonic outflow observed by Parker Solar Probe from 2023-03-16/11:54:58 to 2023-03-17/05:57:20 with unique properties in comparison to typical solar wind.

    The {\alf} and magnetosonic Mach numbers during this period are both $\sim 0.1$ and the sonic Mach number is $\sim 1$. The region shows a massive density depletion, incredibly slow wind speeds ($\sim$150 {\kms}), depletion in the alpha-to-proton abundance ($\sim$0), and slow alpha-to-proton differential speeds. The plasma is magnetically dominated (low $\beta$) and nearly incompressible. The dynamic pressure shows large variation from typical radial dependence, but a fit to the total pressure finds: $\mathrm{P_{expt} = 10^{6.61} \times R^{-3.74}}$ which is a small deviation from the global fit for the entire encounter ($\mathrm{P_{total} \propto R^{-3.71}}$) as $\mathrm{P_{total}}$ is entirely driven by magnetic pressure for this period. 

    \item Following the methods of \citet{Zank-2017, Zank-2020, Zank-2022, Zhao-2022}, we fit the spectra of the {\elas} variables with a nearly incompressible turbulence model and find that this stream is dominated by a slab component ($f^{-3/2}$) unlike the typical $f^{-5/3}$ power law of the 1AU solar wind \citep{Kolmogorov-1941}. Through analysis of one component of noise in the SPAN-I measurements, we caution against making definitive conclusions using the {\zm} spectrum, without deeper analysis of whether it is a physically measurable signal.
    
    
    The stream of interest is highly {\alfic}, characterized by a predominance of outgoing modes with traits akin to large amplitude, spherically polarized waves. Given that the power spectra are dominated by the slab component—the inertial range scaling of {\zp} ($\sim f^{-3/2}$) is consistent with the Nearly Incompressible (NI) turbulence model \citep{Zank-2020, Zank-2022, Zhao-2022}, which predicts scaling exponents of $f^{-5/3}$ and $f^{-3/2}$ for the 2D and slab components, respectively. Simultaneously, the scaling is consistent with MHD turbulence models grounded on the principles of \lq{}critical balance\rq{} \citep{Goldreich-1995, Goldreich-1997} and \lq{}scale-dependent dynamic alignment\rq{} \citep{Boldyrev-2006, Chandran-2015, Mallet-2017}.
    
    In our analysis, we find that the {\zm} spectra flattens around $\mathrm{10^{-2}}$ Hz and attribute this flattening to a variety of noise sources in the SPAN-I measurements rather than a physical mechanism. In Appendix~\ref{sec: appendix-turb}, we outline our quantification of the noise floor associated with the finite velocity grid in the velocity field observations of the SPAN-I instrument (just one source of noise in these measurements) and compare power spectra of pure noise with the spectra of interest. We find that the {\zm} spectral flattening occurs at the lowest level of physically observable fluctuations meaning that turbulent processes at high frequencies are obscured and indicates that this flattening is not due to a physical process and we cannot make conclusive statements about the dissipative range.
    

    \item Observations of this SA stream at {\SO}, Wind, and MMS show the long lived, steady state nature of this outflow and that it is not related to the transient wake of a CME.
    
    Due to multi-spacecraft Parker spiral alignment, we identify the propagation of this stream out to {\SO}, Wind, and MMS. All spacecraft observe (with the expected of {\SO} due to a data gap) the stream close to the expected time of observation based on a Parker streamline model. The stream was observed at 1AU (MMS and Wind) $\sim 8$ days after the observation at Parker, indicating the long lived nature of the stream. Additionally, through basic modeling of CME propagation, we show that the location and time of the observation at 1AU indicate that this stream was not solar wind dynamically controlled by the wake of a CME but rather due to steady state outflow from a single coronal source region.
    
    \item Through PFSS and MHD modeling, we determine that our SA stream outflow originates from a mid-latitude active region.

    Slow wind speeds ($\sim$150 {\kms}), depletion in the alpha-to-proton abundance ($\sim$0), low levels of {\alfty} and alpha-to-proton differential speed are all in situ observations associated with wind streaming from active regions/non-CH origins. Through PFSS and MHD models, we connect this SA stream to a mid-latitude active region due to high relative footpoint brightness, strong photospheric footpoint field strength, and a high magnetic expansion factor \citep{Wang-1990, Riley-2012, Riley-2015}. A comprehensive quantification of errors in the estimated footpoints due to choice of source surface height, error in the SPAN-I velocity measurement used in ballistic propagation, and the ballistic propagation model itself are shown in Appendix~\ref{sec: appendix-model-val} and we find that these error sources do not effect the resulting source region of the SA stream.

    \item We identify a fast wind stream during Parker E15 that propagates from Parker to {\SO} to Wind at 1AU to use as a comparison with our SA stream of interest. Aligned with the findings of \citet{vonSteiger-2000, McComas-1998, McComas-2008} and others, this stream originates from a mid-latitude CH and provides an additional validation method for our modeling results. 

    This fast wind stream shows velocities $\sim$400 {\kms} at Parker, and $\sim$700 {\kms} at {\SO}. From Table~\ref{tab: parameters}, we see that the plasma is highly {\alfic} ({\sigmac} $\sim$1) has high alpha-to-proton abundance, relatively high normalized alpha-to-proton differential velocity, and is compressible -- all characteristics of CH wind. Modeling results reinforce the in situ measurements. The plasma from this period traces to a mid-latitude CH with low footpoint brightness and footpoint magnetic field strength that is below the CH threshold for GONG magnetograms \citep{Stansby-2021aa}. The magnetic expansion factor during this period is small, which for fast wind indicates CH origins \citep{Wang-1990, Riley-2012, Riley-2015} as predicted by the WSA \citep{Arge-2000, Arge-2003, Arge-2004} and \lq{}Distance from a Coronal Hole Boundary\rq{} (DCHB) model \citep{Riley-2001}.

\end{enumerate}

\section{Conclusion} \label{sec: conclusion}


In this study, we discuss the emergence, turbulence properties, and propagation of an 18-hour sub-{\alfic} and near subsonic ({\MS} $\sim$ 1) stream observed by Parker from 2023-03-16/11:54:58 to 2023-03-17/05:57:20. We use PFSS and MHD model approaches to estimate the source region of the observed stream and determine properties of the footpoint. We validate this modeling by checking the alignment of in situ HCS crossings with the modeled neutral line and identifying the source region of a nearby FSW stream to be a CH as expected \citep{vonSteiger-2000, McComas-1998, McComas-2008}. These results align with our expectations in regards to the FSW and HCS solar wind, and in combination with our model validation described in Appendix~\ref{sec: appendix-model-val}, provide strong confidence in our conclusions about the sub-sonic, SA slow solar wind stream.

The combination of in situ observations, $\sim$150 {\kms} speeds, $\sim$0 alpha particle abundance, and high magnetic field strength, coupled with modeling results of high expansion factor and high footpoint field strength shows this stream originated from a mid-latitude active region when compared with the FSW (CH origin) stream. The footpoint field strength is about eight times stronger than that of the CH region and three times as \lq{}bright\rq{} in EUV observations. This relatively high \lq{}brightness\rq{} (intensity) in EUV observations meaning that this wind stream originated from hotter (non-CH) plasma. Multi-spacecraft observations of this low density stream show its long-lived nature and basic CME modeling shows that the unique characteristics of this stream are unlikely to be dynamically controlled by the CME wake as the 1AU observations are well outside the field of influence of the CME per modeling results. As the most extreme and steady SA ({\MA} $\sim$ 0.1) period to date, this shows the capacity for active regions to produce long-lived streams similar to the \lq{}day the solar wind died\rq{} streams \citep[and others]{Usmanov-2000, Usmanov-2005, Stansby-2020c, Halekas-2023}. 

Following the methods of \citet{Zank-2018, Zhao-2022, Zank-2022}, we fit the {\elas} variables with a nearly incompressible (NI) model which combines a 2D MHD component ($f^{-5/3}$) with a slab component ($f^{-3/2}$) to determine the dominant wave mode in the stream. We find that the slab component is the dominant mode unlike in the typical solar wind which follows a $f^{-5/3}$ spectrum. Additionally, we see flattening in the {\zm} spectra at $\sim 10^{-2}$ Hz and discuss this is relation to noise in the measurements from the SPAN-I instrument (see Appendix~\ref{sec: appendix-turb}) meaning we cannot determine the dominant component of the turbulence at small scales. We caution against making conclusive results from {\zm} noting that further work should be done to fully quantify the effects of noise on this measurement.




The work described in this study is limited by the assumptions and implications of the modeling methods and observations available. While we show how errors in the modeling assumptions affect our outcomes, these models still do not provide a complete picture of the state and evolution of plasma in the solar wind. Therefore, this work should be further expanded upon with additional studies delving deeper into the evolution of this stream and perhaps using remote observations from instruments at Earth to look into the composition of the plasma and source region. This stream is seen from Parker out to spacecraft at 1AU and provides a great laboratory to study the effects of propagation on solar wind plasma and differing mechanisms in these density depleted streams in comparison with the ambient wind. Additional studies of the turbulence parameters for this period are important as well, as the sub-{\alfic} region is likely where coronal heating process are generated. 

While studies such as this one, looking at a specific period of slow wind and discussing its source region and characteristics, bring us closer to understanding the true nature and evolution of the solar wind, they are not enough. To fully solve the question of the source region of the slow wind requires a large scale statistical study, ideally with multi-point measurements at varying longitudes and radial distances over differing periods of the solar cycle \citep{Viall-2020}. This calls for a dataset of in situ parameters coupled with remote sensing observations that cover at least a full solar cycle and include particle parameters (velocity, density, temperature) and distribution functions \citep{Wilson-2022wp}, magnetic field measurements (magnetograms, RTN magnetic fields, far side observations), composition metrics (elemental abundance and charge state distributions), and multi-wavelength coverage of coronal outflows \citep{Rivera-2022wp}. This type of dataset requires a new mission concept, one that includes multiple spacecraft and combines the strengths of remote sensing and in situ instruments, as this is the only way to create a full picture of the global structure of the Sun out into the heliosphere.


\section{Acknowledgements} \label{sec: acknowledgements}

The authors would like to thank the reviewer for their comments and suggestions that helped make the manuscript more clear, especially in regards to the discussion of the nearly incompressible model and the flattening of the {\zm} spectra. 

The FIELDS and SWEAP experiments on the PSP spacecraft was designed and developed under NASA contract NNN06AA01C. PR gratefully acknowledges support from NASA (80NSSC20K0695, 80NSSC20K1285, 80NSSC23K0258, and the Parker Solar Probe WISPR contract NNG11EK11I to NRL (under subcontract N00173-19-C-2003 to PSI)).

We acknowledge the NASA Parker Solar Probe Mission and the SWEAP team led by J. Kasper for use of data.

Solar Orbiter is a mission of international cooperation between ESA and NASA, operated by ESA. Funding for SwRI was provided by NASA contract NNG10EK25C. Funding for the University of Michigan was provided through SwRI subcontract A99201MO. 

This work utilizes data produced collaboratively between Air Force Research Laboratory (AFRL) \& the National Solar Observatory (NSO). The ADAPT model development is supported by AFRL. The input data utilized by ADAPT is obtained by NSO/NISP (NSO Integrated Synoptic Program). NSO is operated by the Association of Universities for Research in Astronomy (AURA), Inc., under a cooperative agreement with the National Science Foundation (NSF).

This work utilizes GONG data obtained by the NSO Integrated Synoptic Program, managed by the National Solar Observatory, which is operated by the Association of Universities for Research in Astronomy (AURA), Inc. under a cooperative agreement with the National Science Foundation and with contribution from the National Oceanic and Atmospheric Administration. The GONG network of instruments is hosted by the Big Bear Solar Observatory, High Altitude Observatory, Learmonth Solar Observatory, Udaipur Solar Observatory, Instituto de Astrofísica de Canarias, and Cerro Tololo Interamerican Observatory.

SDO/AIA and SDO/HMI data is courtesy of NASA/SDO and the AIA, EVE, and HMI science teams.

This research used version 4.1.6 of the SunPy open source software package \citep{sunpy}, and made use of HelioPy, a community-developed Python package for space physics \citep{heliopy}. To read MAS results, this work used PsiPy \citep{psipy}. All code to replicate figures can be found at \citet{E15-code}. The authors would additionally like to thank Vamsee Krishna Jagarlamudi for fruitful discussion on the relation of these sub-{\alfic} low density streams with transients.

\software{
\texttt{Astropy} \citep{astropy:2013, astropy:2018, astropy:2022},
\texttt{heliopy} \citep{heliopy},
\texttt{matplotlib} \citep{mpl},
\texttt{numpy} \citep{numpy},
\texttt{pandas} \citep{pandas},
\texttt{pfsspy} \citep{pfss},
\texttt{PsiPy} \citep{psipy},
\texttt{scipy} \citep{scipy},
\texttt{spiceypy}\citep{spiceypy},
\texttt{SunPy} \citep{sunpy}
}

\appendix
\section{Model Validation} \label{sec: appendix-model-val}

\begin{figure} [htb!]
  \includegraphics[width=\columnwidth]{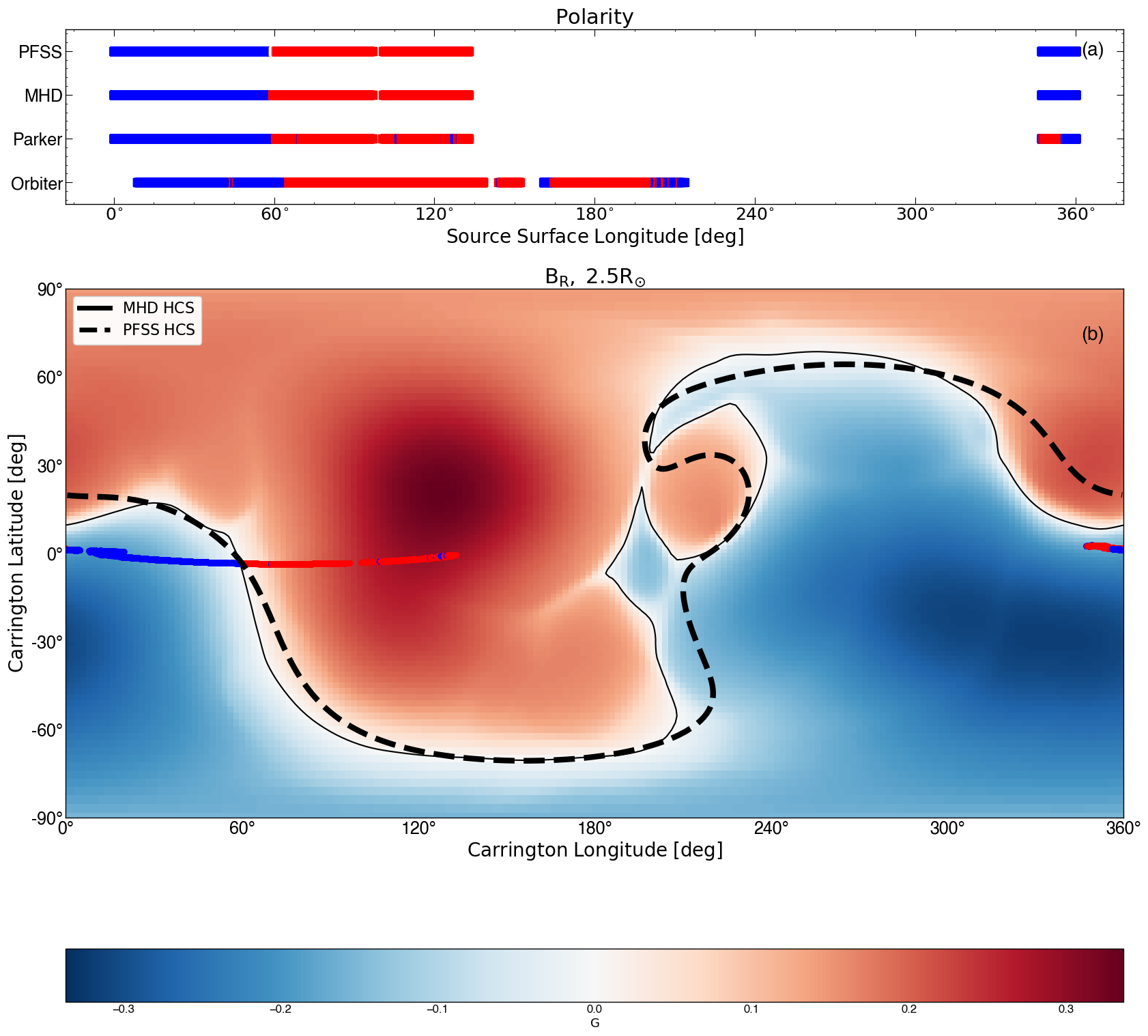}
  \caption{A comparison of model and in situ radial magnetic field polarities. \textit{Panel (a):} A comparison of the polarities measured by Parker and {\SO} with the predicted polarity from the PFSS and MHD solution. \textit{Panel (b):} A radial cut of the MHD solution at the source surface height (2.5{\Rsun}). The modeled HCS from the PFSS and MHD solutions are shown in solid black and dashed black respectively. The Parker trajectory is shown and colored blue (red) for negative (positive) polarity corresponding to the radial magnetic field measured by Parker/FIELDS.}
  \label{fig: mhd_pfss}
\end{figure}

We compare results from the PFSS and MHD models to each other and to in situ data constraints to validate the models and thus our footpoint estimations. In Figure~\ref{fig: mhd_pfss}, we show a comparison of the large scale magnetic field structure between the two models. We see that both the MHD and PFSS solutions produce great alignment with the polarity reversal seen by both {\SO} and Parker. Additionally, the shape of the resulting HCS from the PFSS and MHD solutions show similar characteristics, especially in the region of interest (0{\degree} to 135{\degree}).

\begin{figure} [htb!]
  \includegraphics[width=\columnwidth]{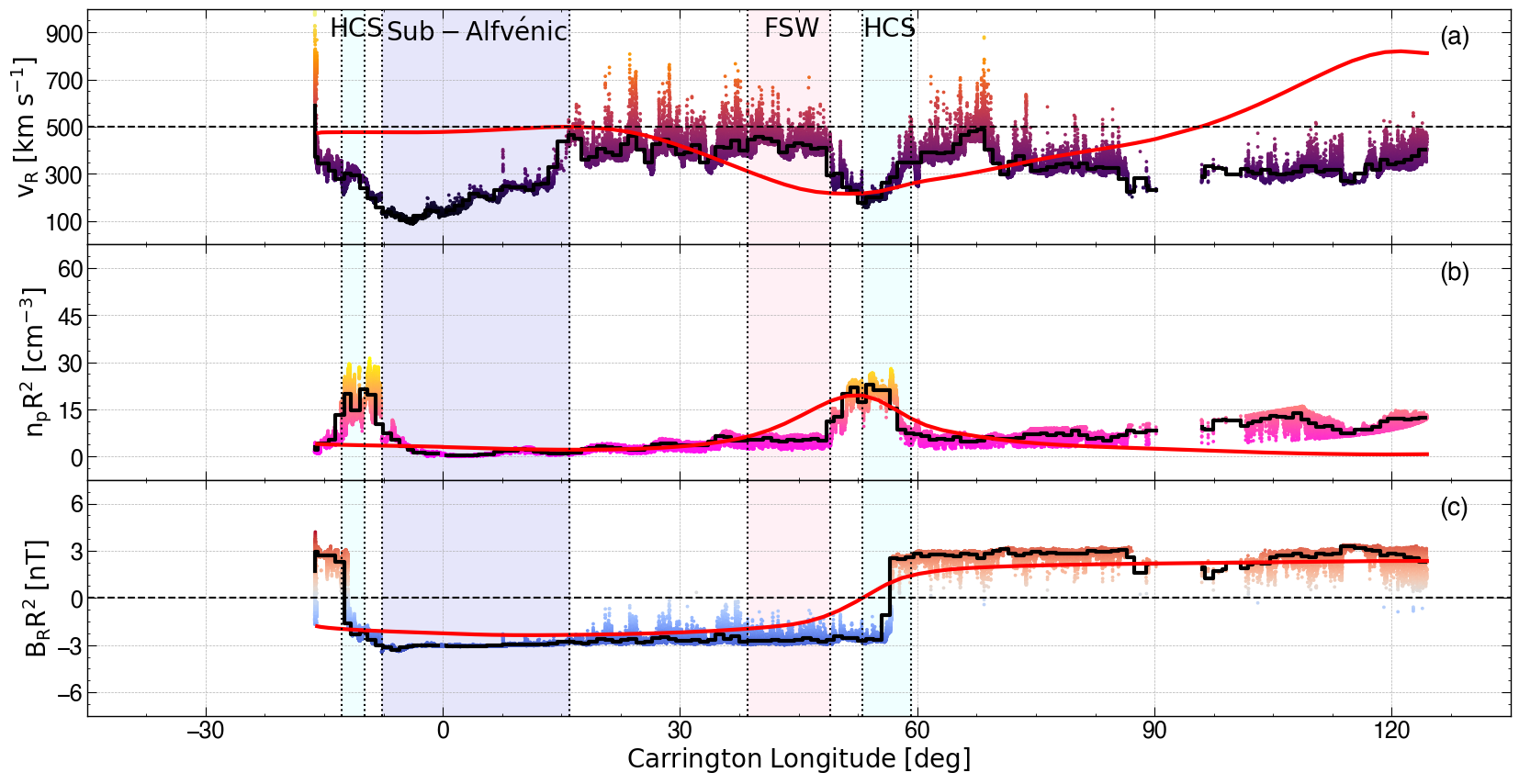}
  \caption{A comparison of observables sampled from the MHD model along the trajectory of Parker (red) in comparison to Parker in situ measurements. We compare the proton radial velocity (panel (a)), scaled number density (panel (b)), and scaled radial magnetic field (panel (c)). In panel (c), we also show the radial position of the spacecraft (AU) in grey. The MHD results are in red. The Parker in situ observations are overlaid with binned data by 1{\degree} in longitude (black).}
  \label{fig: mhd_obs}
\end{figure}

Along with the large scale magnetic field, the MHD solution also produces observables that can be directly compared to in situ measurements by sampling the MHD solution along the trajectory of the spacecraft. In Figure~\ref{fig: mhd_obs}, we compare the proton radial velocity, proton number density, and radial magnetic field values produced by our MHD solution to the in situ measurements from Parker. We are directly comparing the trajectory to the MHD results, which is why instead of plotting the observables as a function of source surface longitude, they are plotted as a function of Carrington longitude (the actual position of the spacecraft). 

We see that for the Parker in situ measurements, the model predicts the structure of the observables near perihelion well, especially around the HCS crossing near 60{\degree} in Carrington longitude. The velocity minimum, density maximum and polarity reversal are shifted by 2{\degree} in longitude off from the in situ measurements. This strong reproduction of the observed velocity, density, and magnetic field structure provides confidence in our modeling and therefore in the footpoint estimations.

\begin{figure} [htb!]
  \includegraphics[width=\columnwidth]{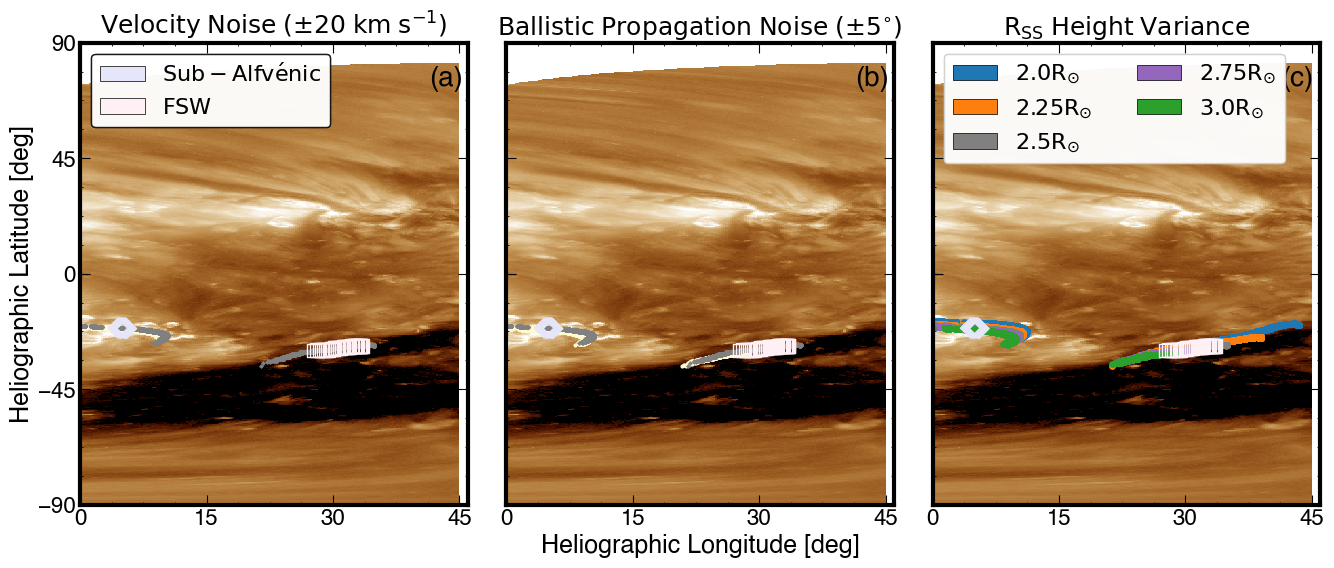}
  \caption{A comparison of how errors in the ballistic propagation and PFSS modeling can effect the resulting footpoints estimations for the wind streams of interest. The \lq{}best\rq{} PFSS model, the one used in this study, is shown in grey in all panels. The SA and FSW footpoints are shown by purple diamonds and pink squares as in Figure~\ref{fig: ar-detection}.
  \textit{Panel (a):} Estimated footpoints with $\pm 20$ {\kms} noise added to the input velocity used for ballistic propagation (pink) compared with the model used (grey).
  \textit{Panel (b):} Estimated footpoints with $\pm 5^{\circ}$ error added to output of the ballistic propagation ballistic propagation (pink) compared with the model used (grey).
  \textit{Panel (c):} Estimated footpoints produced by varying the source surface height of the PFSS model. This study uses the results from {\Rss} = 2.5{\Rsun} shown in grey.
  }
  \label{fig: pfss-fp}
\end{figure}

Figure~\ref{fig: pfss-fp}, shows the effects of errors ($\pm 20$ {\kms}) in the in situ velocity (panel (a)), small ($\pm 5^{\circ}$) errors in the ballistic propagation result (panel (b)), and variance in {\Rss} height (panel (c)) effect the resulting footpoint estimate. Small errors in velocity cause almost non-existent changes to the estimated footpoints (panel (a)), similar to small errors in the ballistic propagation result (panel (b)). They lead to some small changes in the latitude of the footpoint estimation, but do not effect the resulting source region.

We see that the varying the source surface height, has the most dramatic effect on the FSW stream footpoints, expanding the region over which the footpoints cover to include the entire CH as seen in EUV images. This change does not effect the resulting source region as regardless of {\Rss}, the FSW stream comes from a mid-latitude CH. Neither small errors in the observed velocity nor inducing error into the ballistic propagation result effect the resulting source region of the FSW stream.

The same can be said for the resulting source of the SA stream. Variance in the input (velocity) and output of the ballistic propagation do not effect the result that the SA stream emerged from a highly magnetized active region. While varying the source surface height shows some effects on the exact location of the estimated footpoints of the SA stream, it also does not effect the resulting source region of the wind stream.

Through this combination of validation efforts, we show how small errors in ballistic propagation, and variance in the choice of source surface height, do not effect the resulting source region of the streams of interest. Additionally, that both the PFSS and MHD solutions, recreate the global coronal magnetic field in the areas of interest quite well and polarity estimates from the models match with in situ observations. The combination of these factors provides trust in our modeling results, allowing them to compliment the observations in discussion and characterization in the source regions of the streams of interest.

\section{Discussion of Uncertainties in the Calculation of the {\elas} Variables} \label{sec: appendix-turb}

In Section~\ref{sec: turb-spectra} we outlined our methods for calculating the PSD of the turbulence parameters and fitting the {\elas} variables using a nearly incompressible (NI) model which combines 2D MHD turbulence with a slab term. In Figure~\ref{fig: turbulence_psd} we see a flattening of the {\zm} spectra, which also has been previously observed by \citet{Zhao-2022, Zank-2022}. \citet{Zhao-2022} suggests this may be due to physical processes while \citet{Zank-2022} argues that it has to be due to the noise floor of the SPAN-I instrument. We agree with \citet{Zank-2022} as our analysis suggests this flattening to be partially due to the finite velocity grid of SPAN-I, along with other sources of error that we do not quantify in this study. This analysis does not include other instrumental sources of error and more should be done to fully quantify the effect of noise from SPAN-I on our ability to precisely measure the {\elas} variable spectra. Rather this serves as a caveat to the {\zm} spectra shown here and to spectra shown in other studies to say that more in depth analysis must be done to determine whether any science can be done with this measurements, especially in the high frequency regime.

\begin{figure} [htb!]
  \includegraphics[width=\columnwidth]{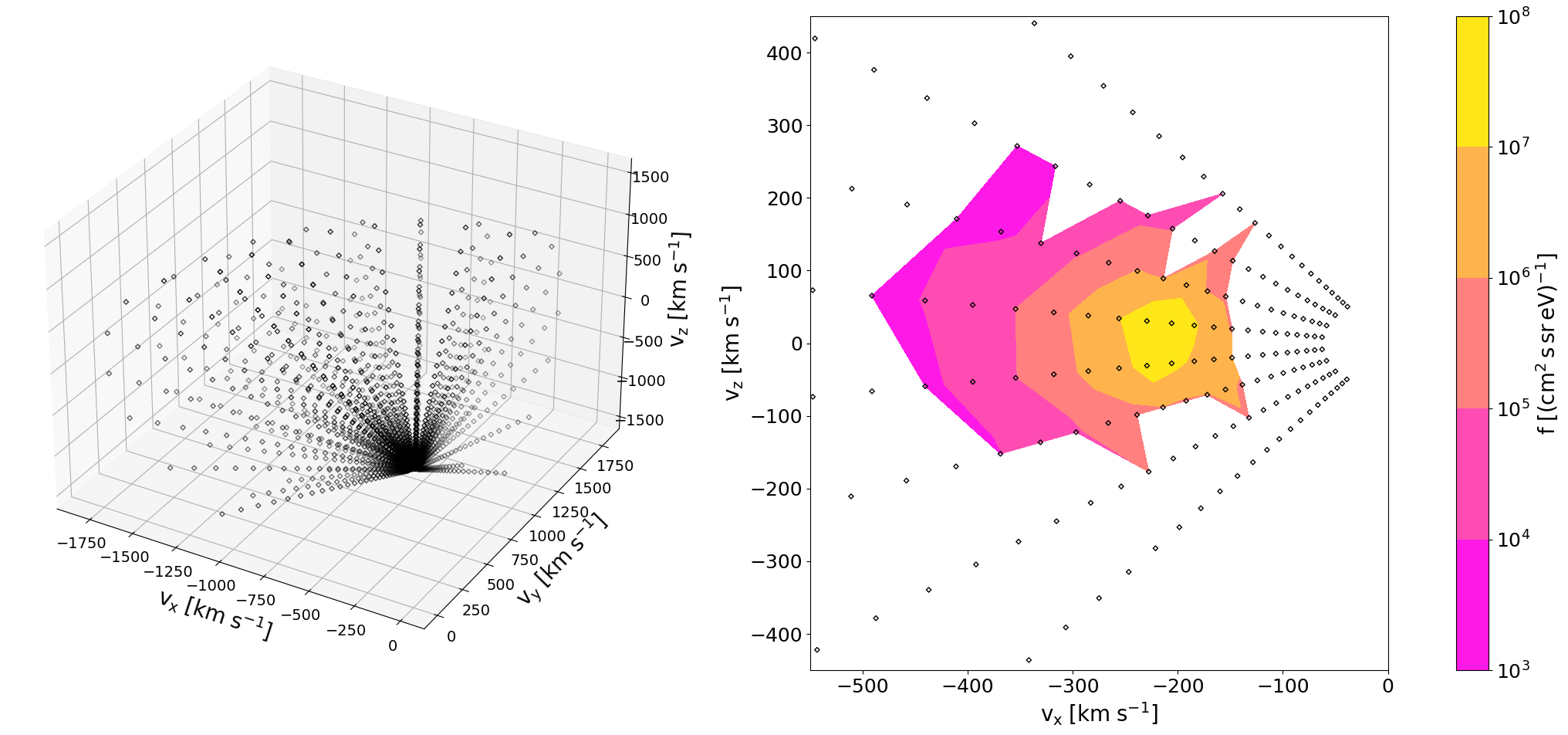}
  \caption{An example of the finite velocity grid of SPAN-I and VDF for the SA stream (March 16, 2023 14:00). Left panel: Non-uniform 3D velocity grid showing the points over which a VDF is measured. Right panel: VDF contours summed and collapsed onto the $\theta$ plane.}
  \label{fig: span-energy}
\end{figure}

We first seek to quantify the smallest measurable fluctuations due to the finite grid of SPAN-I, noting that this study does not consider the other sources of noise in this measurement. The {\dv} fluctuations, $\mathbf{v} - \mathbf{v_0}$ where $\mathbf{v_0}$ is the background velocity field over a running 20-minute time window, are dependent on the smallest measurable fluctuation possible by the SPAN-I instrument. The Level 3 SPAN-I protons density, velocity, and temperature parameters used to calculate the turbulence parameters ({\dv}, {\db} and {\elas} variables) are the 3D moments of the SPAN-I proton velocity distribution functions or VDFs \citep{Livi-2022}. Using the SPAN-I Level 2 VDFs, we convert the $\theta$, $\phi$, and energy spectra to velocity measurements with the following relations:
\begin{align*}
    v = \sqrt{2 e \mathrm{E}/m_p} \\
    v_x = v \cos{\phi} \cos{\theta} \\
    v_y = v \sin{\phi} \cos{\theta} \\
    v_z = v \sin{\theta} \\
\end{align*}
This gives SPAN-I energy bins in velocity space. VDFs are determined over this finite velocity grid and an example of the velocity grid and an associated VDF for the SA stream is shown in Figure~\ref{fig: span-energy}. 

There is a lower limit to the velocity field fluctuation that can be measured due to the spacing between the velocity bins (Figure~\ref{fig: span-energy}) alongside other instrumental sources of error (not discussed here). We can quantify the minimum measurable fluctuation due to the effect of the finite velocity grid using the mean velocity for our time period and the spacing between bins with the following steps:
\begin{enumerate}
    \item Fit the VDF to find the component wise average velocity over our time period, we call this velocity $\mathbf{v}_{mom}$. This can be compared to the moment fits from the Level 3 SPAN-I measurements for the same period.
    \item Identify the neighboring cube of the average velocity found in step one ($\mathbf{v}_{mom}$). These are the eight adjacent bins ($\mathbf{v}_i$) to $\mathbf{v}_{mom}$ in velocity space as seen in the left panel of Figure~\ref{fig: span-energy}.
    \item Find the distance from $\mathbf{v}_{mom}$ to each neighboring point: $|\mathbf{v}_{mom} - \mathbf{v}_i|$ for $i=1, ..., 8$.
    \item Average these distances, $\frac{\sum_{i=1}^8 |\mathbf{v}_{mom} - \mathbf{v}_i|}{8}$, to quantify the smallest measurable velocity fluctuation from SPAN-I during the period.
\end{enumerate}

Using this method on the VDF shown in Figure~\ref{fig: span-energy}, we find $\mathbf{v}_{mom}$ is (-193, 106, 43) {\kms} with a magnitude of 225 {\kms}. We then place this point on the velocity grid shown in Figure~\ref{fig: span-energy} and find the neighboring eight points. From there, we take the Euclidean distance and average over the eight points to get a value of $\sim 5$ {\kms}. 

We do this same analysis over the three intervals that were used to calculate the spectra in Figure~\ref{fig: turbulence_psd}. The results are shown in Table~\ref{tab: noise-floor}.

\begin{table}
  \centering
  \begin{tabular}{|c||c|c|c|c|}
    \hline 
    \textbf{Time Period} & \textbf{$\mathbf{v}_{mom}$} & \textbf{Expected Flattening} \\
    \hline
    \hline
    \textbf{Interval One: 2023-03-16/12:00 - 18:00} & (-193, 106, 43) & 5.1 {\kms}  \\
    \hline
    \textbf{Interval Two: 2023-03-16/18:00 - 23:59} & (-238, 120, 49) & 19.2 {\kms} \\
    \hline
    \textbf{Interval Three: 2023-03-17/00:00 - 06:00} & (-317, 172, 29) & 60.5 {\kms}  \\
    \hline
  \end{tabular}
  \caption{Overview of the parameters and results from quantifying the smallest measurable  fluctuation due to the finite velocity grid of the SPAN-I observations.}
  \label{tab: noise-floor}
\end{table}



\begin{figure} [htb!]
  \includegraphics[width=\columnwidth]{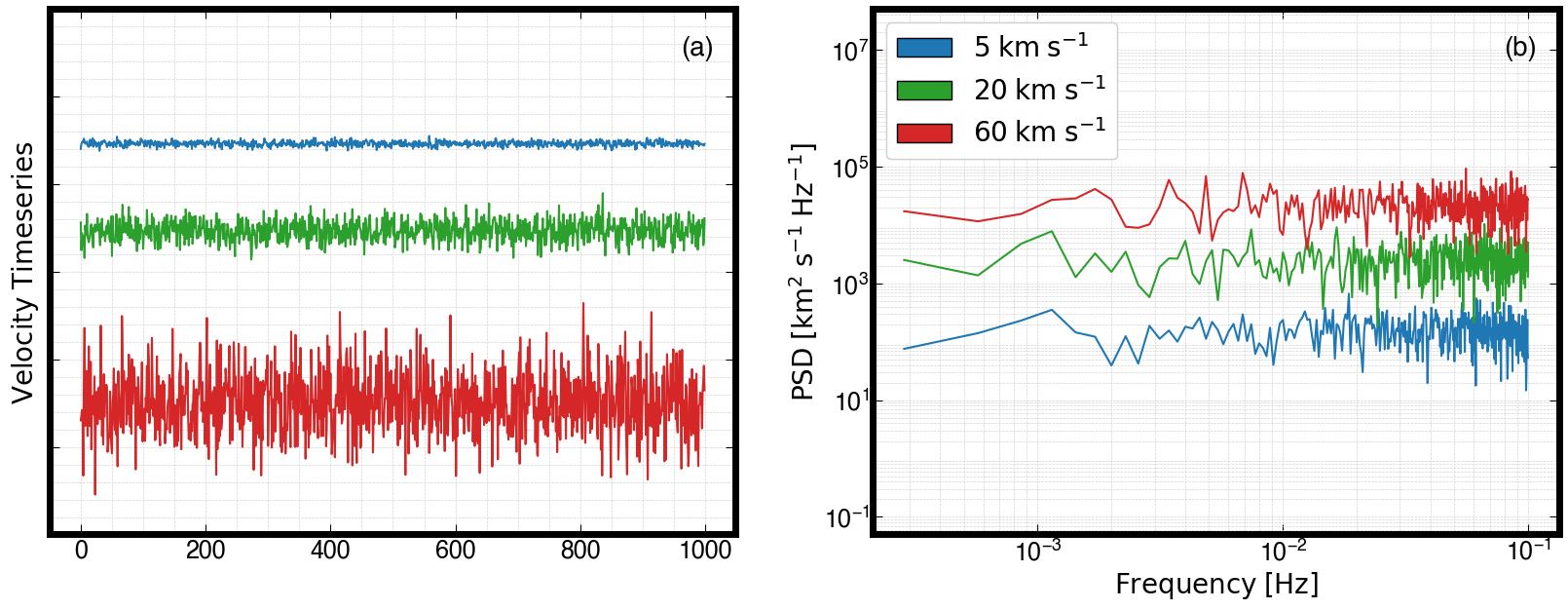}
  \caption{A comparison of the PSD for a velocity timeseries (200 {\kms}) with varying noise levels ({\dv}). Time series of 200 {\kms} speed with varying levels of random noise applied: 5 {\kms} (blue), 10 {\kms} (green), and 60 {\kms} (red). The offset between each timeseries for visualization purposes. \textit{Panel (b):} PSD of each of the velocity timeseries in panel (a) colored in the same manner.}
  \label{fig: noise-psd}
\end{figure}

The next step is to directly compare what a pure \lq{}noise\rq{} {\dv} measurement would look like in comparison to the actual fluctuation and {\elas} variable spectra. Following the PSD method in Section~\ref{sec: turb-spectra}, we calculate the power spectra for a fake timeseries of speeds at 200 {\kms} with varying levels of noise corresponding to {\dv}. These time series are created at the same cadence as the observations used for the power spectra in Figure~\ref{fig: turbulence_psd}. Examples of these are shown in Figure~\ref{fig: noise-psd}.

\begin{figure} [htb!]
  \includegraphics[width=\columnwidth]{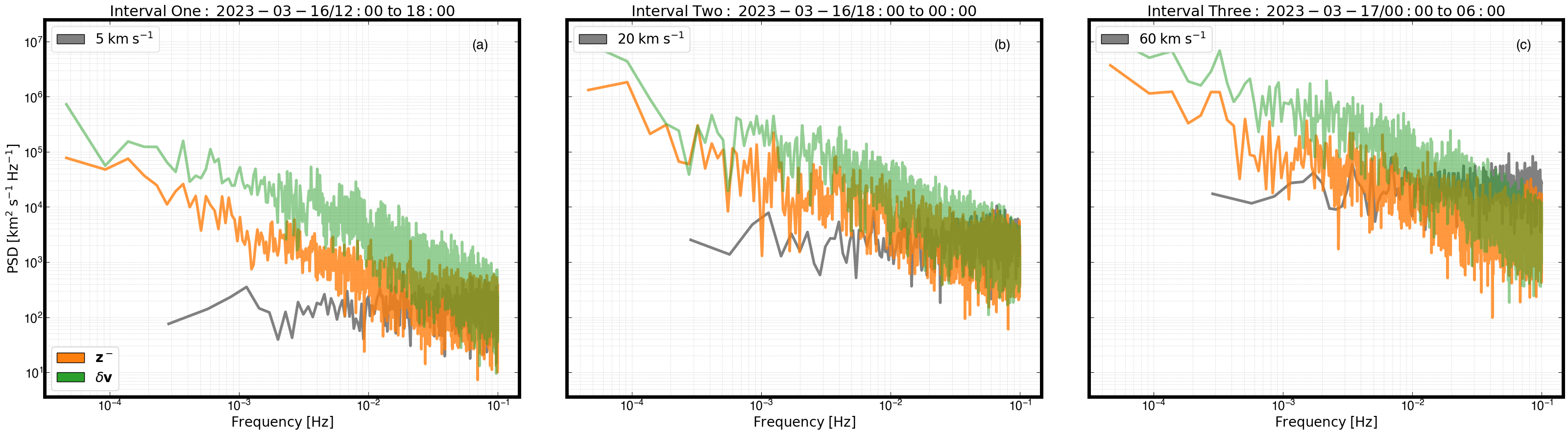}
  \caption{A comparison of the PSD for varying noise levels (5, 20, and 60 {\kms}) with the computed PSD for the {\zm} variable from Figure~\ref{fig: turbulence_psd} for each of the three intervals. Each panel shows the PSD of each of the velocity timeseries from Figure~\ref{fig: noise-psd} colored in the same manner.}
  \label{fig: noise-psd-spec}
\end{figure}

In Figure~\ref{fig: noise-psd-spec}, we compare our computed spectra with the spectra of the different {\dv} levels, to approximate the lowest level of measurable velocity fluctuations in our calculation. We do this separately for the three intervals outlined in Section~\ref{sec: turb-params} and compare this with the lowest level we expect to be measurable based on the description above. 


For each interval, we compare the level at which the spectra does (or does not) flatten with the expectation based on the quantification of the measurable fluctuations due to the finite velocity grid as described above. In interval one, the {\zm} spectra flattens at $\sim$ 5 {\kms}, as expected. For interval two, we see that the {\zm} does not flatten as significantly as the other two periods, but shows signs of flattening $\sim 20$ {\kms}. In the third interval, flattening occurs a bit below $\sim 60$ {\kms}, likely due to a combination of noise from the finite grid and other instrumental sources. We would expect intervals two and three to flatten at higher amplitudes than interval one due to the larger fluctuation amplitudes of the turbulence parameters during this period (Figure~\ref{fig: turbulence}).

It is important to note that there could be periods in which the {\dv} spectra do not flatten, however this is not a sign that the {\zm} flattening is physical. Since {\zm} is measure of the difference in fluctuations ($\mathbf{z}^{-} = \delta \mathbf{v} - \delta \mathbf{b}$) we are often measuring a difference that is smaller than the noise in the velocity measurements. This is the reasoning for not being able to trust the {\zm} measurements at high frequencies and why making conclusions off the {\zm} measurement in general must be done carefully.

Therefore, we believe this flattening to be due to the finite velocity grid and other effects of instrumental noise that we did not account for limiting the physically measurable fluctuations rather than physical effects. This analysis shows a caveat to understanding the high frequency regime using SPAN-I observations and similar analyses should be used for further studies of the turbulence properties observed at Parker Solar Probe. 

Additionally, we can look at these and other results from a mathematical perspective. Specifically we are interested in the fact that the {\dv} spectrum during this period drops below the {\zm} spectra. Starting with the definitions of the fluctuations and {\elas} variables, we have:

\begin{align*}
    \delta \mathbf{v} = \mathbf{v} - \mathbf{v_0} \\
    \delta \mathbf{b} = \mathbf{b} - \mathbf{b_0} \\
    \mathbf{z}^{\pm} = \delta \mathbf{v} \pm \delta \mathbf{b}.
\end{align*}

Taking the power of {\zp} and {\zm} we find the following:
\begin{align*}
     \mathbf{z}_+^2 = \delta \mathbf{v}^2 + \delta \mathbf{b}^2 + 2 \delta \mathbf{v} \cdot \delta \mathbf{b} \\
     \mathbf{z}_-^2 = \delta \mathbf{v}^2 + \delta \mathbf{b}^2 - 2 \delta \mathbf{v} \cdot \delta \mathbf{b}
\end{align*}


In order for $\mathbf{z}_-^2 >$ {\dv}$^2$ then based on the above, $\delta \mathbf{b}^2 > 2 \delta \mathbf{v} \cdot \delta \mathbf{b} = (\mathbf{z}_+^2 - \mathbf{z}_-^2) / 2$ meaning the $2 \delta \mathbf{v} \cdot \delta \mathbf{b}$ cross term is either zero or negative. $2 \delta \mathbf{v} \cdot \delta \mathbf{b}$ goes to zero when {\dv} and {\db} are decorrelated, and in the case where {\dv} is very small, then $\mathbf{z}_-^2 \sim$ {\db}$^2$ and $\mathbf{z}_-^2$ $>$ {\dv}$^2$. $\mathbf{z}_-^2 \sim$ {\db}$^2$ is seen in all the observational spectra shown in this paper. 

We then check whether {\dv} is all noise. We can write the {\zp} and {\zm} power spectra as a function of {\db} and {\dv} in the Fourier domain. As in Section~\ref{sec: turb-spectra}, we define $\tilde{x}$ as the Fourier transform of timeseries $x$. Then we can say:

\begin{equation}
    \tilde{\mathbf{z}}^{\pm} = \tilde{v}\tilde{v}^* + \tilde{b}\tilde{b}^* \pm (\tilde{v}\tilde{b}^* - \tilde{v}^*\tilde{b})
\end{equation}

where the $\tilde{v}\tilde{b}^* - \tilde{v}^*\tilde{b}$ term is the cross term in Fourier space. We find that the coherence of the cross term is small at high frequencies for both \citet{Zhao-2022} SA periods, and for the one of interest in this study.

Therefore, $\mathbf{z}_-^2 > \delta \mathbf{v}^2$ is expected when {\dv} is noise ({\dv} $= \epsilon \mathbf{v}$). Since the {\zm} spectra at the point of flattening reach a noise threshold of 3 {\kms}, below the noise floor of the SPAN-I instrument during this period, we assume the velocity measurement is entirely noise ($\epsilon$). This suggests that {\zm} $= \epsilon \mathbf{v} - \delta \mathbf{b}$ is not a physical measure of the {\zm} spectrum.



\bibliography{ms}{}
\bibliographystyle{aasjournal}

\end{document}